\documentclass[10pt]{article}
\usepackage{color}
\usepackage{amssymb}
\usepackage{amsbsy}
\usepackage[colorlinks=true, pdfstartview=FitV, linkcolor=blue, citecolor=blue, urlcolor=blue]{hyperref}
\usepackage{verbatim}
\usepackage{epsfig}
\def\ds{\displaystyle}

\def\res{\mathop{\mathrm{res}}\limits_}

\newtheorem{theorem}{Theorem}[section]
\newtheorem{remark}{Remark}[section]
\newtheorem{proposition}{Proposition}[section]
\newtheorem{corollary}{Corollary}[section]
\newtheorem{example}{Example}[section]
\newtheorem{lemma}{Lemma}[section]
\newtheorem{definition}{Definition}[section]
\makeatletter
\@addtoreset{equation}{section}
\makeatother
\def\tr{\mathrm {Tr}}
\def\le{\left}
\def\bc{\begin{corollary}}
\def\ec{\end{corollary}}

\def\ri{\right}
\def\br{\begin{remark}\rm\small}
\def\1{{\bf 1}}
\def\er{\end{remark}}
\def\bt{\begin{theorem}}

\def\et{\end{theorem}}

\def\bx{\begin{example}}
\def\ex{\end{example}}
\def\bd{\begin{definition}}
\def\ed{\end{definition}}
\def\bp{\begin{proposition}\rm}
\def\bl{\begin{lemma}\em}
\def\el{\end{lemma}}
\def\ep{\end{proposition}}
\def\be{\begin{equation}}
\def\ee{\end{equation}}
\def\bea{\begin{eqnarray}}
\def\eea{\end{eqnarray}}
\def\beaq{\begin{eqnarray}}
\def\eeaq{\end{eqnarray}}
\def \pa{\partial}
\def\C{{\mathbb C}}

\begin{document}
\def\L{\Lambda}
\begin{flushright}
\end{flushright}
\vspace{0.2cm}
\begin{center}
\begin{Large}
\textbf{Isomonodromic deformation of resonant rational connections}
\end{Large}\\
\vspace{1.0cm}
M. Bertola$^{\dagger\sharp}$\footnote{Work supported in part by the Natural
    Sciences and Engineering Research Council of Canada (NSERC),
    Grant. No. 261229-03 and by the Fonds FCAR du
    Qu\'ebec no. 88353.}\footnote{e-mail:
  bertola@mathstat.concordia.ca}, M. Y. Mo$^{\sharp}$
\footnote{e-mail: mo@crm.umontreal.ca}
\bigskip\\
\begin{small}
$^{\dagger}$ {\em Department of Mathematics and
Statistics, Concordia University\\ 7141 Sherbrooke W., Montr\'eal, Qu\'ebec,
Canada H4B 1R6}  \\
\smallskip
$^{\sharp}$ {\em Centre de recherches math\'ematiques,
Universit\'e de Montr\'eal\\ C.~P.~6128, succ. centre ville, Montr\'eal,
Qu\'ebec, Canada H3C 3J7} \\
\end{small}
\bigskip
{\bf Abstract}
\end{center}
We analyze  isomonodromic deformations of rational connections on
the Riemann sphere with Fuchsian and irregular singularities. The
Fuchsian singularities are allowed to be of arbitrary resonant index;
the irregular singularities are also allowed to be resonant in the
sense that the leading coefficient matrix at each singularity may
have arbitrary Jordan canonical form, with a genericity condition
on the Lidskii submatrix of the subleading term.
We also give the relevant notion of isomonodromic tau function extending the one of
non-resonant deformations introduced by Miwa-Jimbo-Ueno. The tau
function is expressed purely in terms of spectral invariants of the
matrix of the connection.

\tableofcontents
\section{Introduction}
The history of isomonodromic deformations
dates back to Schlesinger \cite{schlesinger} who studied rational
connections of size $r\times r$  with simple poles (Fuchsian singularities)
\be
A(x) =\sum_{\gamma_i} \frac {A_i}{x-\gamma_i}
\ee
and defined a system of differential equations  describing the dependence of the
residue matrices $A_i$ on the position of the poles $\gamma_i$ under
the condition that the monodromy representation induced by the kernel
of the connection is independent of the position of the poles.

Later on the authors of \cite{JMU} described an extended system of
equations where {\em irregular} singularities were considered;
\be
A(x) =\sum_{\gamma_i} \sum_{k=0}^{r_i}\frac {A_{i.k}}{(x-\gamma_i)^{k+1}}
\ee
 in this
case the differential equations are with respect to certain exponents
in the formal asymptotic behavior of the kernel solution and the
``monodromic'' data are extended to include the parameters appearing in
Stokes' phenomenon.

In both settings the study was limited to {\bf non-resonant}
connections. The notion of resonance for a rational connection is as follows; for a
Fuchsian singularity we say that the connection is resonant if the
residual spectrum at a singularity (the spectrum of the residue matrix) contains
eigenvalues differing by a nonzero integer. An irregular
singularity  is called resonant if the eigenvalues of the
leading coefficient are repeated.

In the past years significant classes of examples have appeared in
which one is forced to consider resonant isomonodromic
deformations. Probably one of the most relevant is in the analysis of
quantum cohomology using the notion of Frobenius manifolds
\cite{dubrovin}, where a Fuchsian resonant singularity appears in the
relevant isomonodromic deformation.

Lately, the analysis of the Riemann--Hilbert problem  associated to
biorthogonal polynomials for the multi-matrix models has lead to a
system with an irregular resonant singularity \cite{BEH, BEH2}.

The general structure of an isomonodromic deformation of a resonant Fuchsian
connection was addressed in \cite{bolibruch} whereas it seems
that no attempt is being made in the literature to address
isomonodromic deformations of irregular resonant singularities in general.
It is the purpose of this paper to analyze these issues.
The completely general classification of Fuchsian resonant
singularities and their isomonodromic deformations are essentially
contained in \cite{bolibruch} and we are rephrasing it in Section
\ref{se:fuch} for the reader's convenience. The main
feature which distinguishes these deformations from the usual
Schlesinger equations is that in  a monodromy-preserving deformation of a resonant
Fuchsian connection the deformation matrices may in fact have higher
order poles, of degree at most equal to the maximal integral
difference between two residual eigenvalues at each pole (the
{\bf resonance index}); these higher order poles -however- are the result of
the bigger ``local'' gauge freedom that arises due to the resonant
character of the singularity and can in fact always be gauged to zero
by a rational gauge equivalence (without changing the position of the
poles and  the Fuchsian
character of the connection, see Sect. \ref{se:fuch}).

The situation is not dissimilar from the nonresonant case, in which
-however- the gauge freedom is restricted to a point where the only
arbitrariness is  global constant gauge transformations.  \par\vskip
5pt

The case of isomonodromic deformations
of irregular resonant singularities  (in the generalized sense of
\cite{JMU}) is quite unexplored, mainly because
of the  difficulty in analyzing the normal asymptotic form near any such
singularity.

The class of resonant singularities which we analyze here may well be
considered ``minimally'' resonant in the sense to follow:
 we will consider rational connections $A(x)$ such that
near an irregular singularity $x=\gamma$ the leading coefficient
matrix $A_{r,\gamma}$  may have
an arbitrary Jordan canonical form. However we impose a
(completely explicit) genericity assumption on the second-leading coefficient matrix as
explained in Section \ref{se:Lidskii}.

The meaning of our genericity assumption has a clear interpretation
in terms of the spectral curve of the connection $A(x)$, i.e. the
algebraic curve satisfied by the eigenvalues $y(x)$ of $A(x)$.
 It ensures that  the branching structure of the (desingularized)
spectral curve above $x=\gamma$ is in agreement with the  dimension
 of the Jordan cells of the leading coefficient matrix.

This allows us to obtain a canonical form for the asymptotic behavior
of a kernel solution  $\Psi' = A \Psi$ (Thm. \ref{main}), which is the
crucial tool in order to construct isomonodromic deformations and
prove their compatibility.

\paragraph{Tau Function.} Another main point of our construction is
that we define -following \cite{JMU}- the notion of ``isomonodromic
tau function'' for the deformations we have defined. Here our approach
differs radically from that in \cite{JMU} inasmuch as our definition
of tau function is not obtained in terms of the formal asymptotic
data. We rather use the spectral curve itself, thus showing explicitly
the spectral nature of the tau function. The approach is along the
same lines of \cite {BHHP}, where it was shown that Miwa-Jimbo-Ueno's
tau function is in fact a spectral invariant expression.

Let us briefly comment on the necessity and interest in a definition
of tau function in this context; it has been shown repeatedly
\cite{BEHiso, BGek,ITW} that JMU's tau function coincides with
T\"oplitz/H\"ankel determinants of moments of
measures. This allows to establish a direct connection between the
partition function of certain matrix models and the tau function of a
(naturally) associated isomonodromic deformation.

In the context of multimatrix models we have many of the same
features, namely a partition function and isomonodromic deformations
of a certain ODE \cite{BEH,BEH2} and it is natural to imagine an
analogous relationship with an isomonodromic tau-function. The main
obstacle is the absence of a general definition of tau function for
resonant irregular singular ODEs.

Although the resonance of the ODE appearing in the two-matrix model is
not quite of the class which we analyze in this paper, on the other
hand it has many of the same features. In particular the tools
developed here are sufficient to analyze that situation (although in a
{\em ad hoc} way). This analysis is contained in a separate paper.

\section{Preamble: perturbations of spectra and Lidskii coefficients}
\label{se:Lidskii}
In this section we introduce some necessary notations and definitions
that will be used in the analysis of the formal asymptotics of
singular ODEs. The philosophy inspiring these considerations is that
solving a singular ODE by formal series is -to high order- the "same"
as finding perturbative  eigenvectors of an analytic (formal)
perturbation. We recall some relevant facts about analytic (or
formal) perturbations of the spectrum of matrices. From our point of
view the issue of convergence of the series that will appear is
irrelevant since  that in our applications only a finite number of the
coefficients will appear; for this reason we will limit the discussion
to the formal aspects of the problem, with the understanding that
under mild additional assumptions the considerations to come could be
set in an analytic framework.\par
By {\em perturbation} we mean a (formal) power series in a small parameter $\epsilon$ of the form
\be
M(\epsilon):= M_0 +\epsilon M_1 + \mathcal O(\epsilon^2)\ .
\ee
One of the main question in perturbation theory is to understand the
behavior of the spectrum of $M(\epsilon)$ and its relation to the
"unperturbed" spectrum of $M(0)$.  A related question is that of
describing the perturbation of the corresponding eigenvectors. \\
The generic case  is very well understood and studied and corresponds
to the case where all unperturbed eigenvalues are distinct and simple
(i.e. with algebraic multiplicity one). In this case it is not hard to
show that each perturbed  eigenvalue admits a power series expansion
(possibly formal) \cite{kato}
\be
\lambda_j(\epsilon) = \lambda_j(0) + \mathcal O(\epsilon).
\ee
Complications arise when the unperturbed spectrum consists of
eigenvalues with algebraic multiplicity higher than one, namely when
the Jordan canonical form of $M_0$ is allowed to be the most
general. Without loss of generality we may assume that $M_0$ is in its
Jordan canonical form. Adopting Arnold's \cite{arnold} notation we
denote a matrix in Jordan canonical form by the product of the
determinants of its blocks. For example $\alpha^3 \alpha^2\beta ^4$
denotes a matrix with two Jordan blocks with eigenvalue $\alpha$ of
dimension $3\times 3$ and $2\times 2$, and another Jordan block of
size $4\times 4$ with eigenvalue $\beta$. In this example then the
algebraic multiplicity of $\alpha$ is $3+2=5$ but the geometrical
multiplicity (the rank of the eigenspace) is $2$.

As a general rule a multiple eigenvalue like the $\alpha$ in the
example, will split under perturbation in $5$ distinct eigenvalues; in
fact under certain genericity assumption on the first jet of the
perturbation (i.e. $M_1$) the splitting that will occur is in a
triplet and a doublet according to the size of the two Jordan
blocks. The eigenvalues of the triplet will be Puiseux series in
$\epsilon^{\frac 1 3}$ and will be cyclically permuted after a loop
around the origin in the $\epsilon$ plane; similarly the eigenvalues
of the doublet will be Puiseux series in $\epsilon^{1/2}$ enjoying a
similar cyclicity.  \par
In \cite{lidskii} were studied sufficient conditions for the splitting
of an eigenvalue to occur in  cyclic $k$-tuplets according to the
sizes of the Jordan blocks of the unperturbed matrix $M_0$. More
recently the approach of Lidskii has been refined (see
\cite{jeannerod} and references therein) to handle the cases in which
the splitting of eigenvalues does not necessarily respect the Jordan
decomposition of $M_0$.
For the purposes of the present paper we restrict ourselves to the "generic" stratum within this class.
 \par
 In order to describe this theory we start by observing that we can
 limit ourselves to the case where $M_0$ has a single eigenvalue which
 we can set to zero by shifting with the identity matrix. Indeed
 blocks with distinct eigenvalues will have eigenvalues which will not
 "mix" under a small perturbation.
\\
Suppose thus that $M_0 = 0^{n_1} 0^{n_2}\cdots 0^{n_K}$; we arrange these blocks in weakly decreasing order
\be
n_1\geq n_2 \geq \dots  \geq n_K.
\ee
We next partition the first jet of the perturbation ($M_1$)  in $K\times K$ blocks
according to the same block decomposition of $M_0$; from each of these
blocks of $M_1$ we extract the lower-left entry and form a $K\times K$
matrix; we will call such matrix $L(M_1)_{\{n_1,n_2,\dots,n_K\}}$ the
{\bf Lidskii matrix} of $M_1$ subordinated to the block decomposition
of $M_0$.

\bea
M_0 =\left[\begin{array}{cccc|cc|cc|c}0 & 1 & 0 & 0 & 0 & 0 & 0 & 0 &
    0 \\0 & 0 & 1 & 0 & 0 & 0 & 0 & 0 & 0 \\0 & 0 & 0 & 1 & 0 & 0 & 0
    & 0 & 0 \\
0 & 0 & 0 & 0 & 0 & 0 & 0 & 0 & 0 \\
\hline
0 & 0 & 0 & 0 & 0 & 1 & 0 & 0 & 0 \\
0 & 0 & 0 & 0 & 0 & 0 & 0 & 0 & 0 \\
\hline
0 & 0 & 0 & 0 & 0 & 0 & 0 & 1 & 0 \\
0 & 0 & 0 & 0 & 0 & 0 & 0 & 0 & 0 \\
\hline
0 & 0 & 0 & 0 & 0 & 0 & 0 & 0 & 0\end{array}\right]\\
M_1 = \left[\begin{array}{cccc|cc|cc|c}
 && & & &   &   &   & \\
 &   &   &   &   &   &   &  &  \\
  &    &   &   &   &   &   &   &   \\
  L_{11} & &  &   &L_{12} & & L_{13} &  & L_{14} \\
\hline
   & &   &   &  &   &  &   &   \\
   L_{21} &&   &   & L_{22} & & L_{23} & & L_{24} \\
\hline
     &   &   &   &   &   &   &   &   \\
     L_{31} &   &   &   & L_{32} &   & L_{33} &   & L_{34} \\
\hline
     L_{41} &   &   &  & L_{42} && L_{43} &  & L_{44}
  \end{array}\right]
\eea
 Let $\ell_1$ be the number of blocks of the same size as $n_1$: the
 next block of strictly smaller dimension will be then the
 $\ell_1+1$. Let $\ell_2$ be the number of blocks of the  same size as
 the $n_{\ell_1+1}$, and so on and so forth.
 At the end of this procedure the $K$ diagonal blocks are grouped together according to the dimensions
 \be
\underbrace{ n_1,n_2,\dots ,n_{\ell_1}}_{\ell_1}, \underbrace{n_{\ell_1+1}, \dots, n_{\ell_1+\ell_2}}_{\ell_2},\dots\ .
 \ee
 This grouping induces a partitioning of the Lidskii matrix $L$ into
 blocks; for the example in the figure above we have
 ($\ell_1=1,\ell_2=2,\ell_3=1$)
 \be
 L = \left[\begin{array}{c|cc|c}L_{11} & L_{12} & L_{13} & L_{14} \\\hline L_{21} & L_{22} & L_{23} & L_{24} \\
 L_{31} & L_{32} & L_{33} & L_{34} \\ \hline L_{41} & L_{42} & L_{43} & L_{44}\end{array}\right]
 \ee
We now consider the principal block-submatrix of $L$ according to this
 decomposition: namely the first principal block submatrix is of size
 $\ell_1\times \ell_1$, the next is of size $(\ell_1+\ell_2)^2$, etc.
For each of these submatrices we construct the pseudo-characteristic
 polynomial, namely the determinant of the submatrix minus $\lambda$
 times the projector onto the lower right sub-matrix. At each step we
 have a submatrix of size $(\ell_1+\dots + \ell_j)^2$ and the
 corresponding pseudo charpoly is of degree $\ell_j$.
\bd\label{de:generic}
The roots of these polynomials will be called the {\bf Lidskii pseudovalues}.\\
We will say that they are {\bf generic} if none of them is zero and the discriminant of each pseudocharpoly is nonzero.
\ed

We think that the description of the procedure is sufficiently
involved to require an example: continuing with the above one, the
Lidskii pseudo-charpoly's are
\bea
&&P_1 = \det[L_{11}-\lambda]\\
&&P_2 = \det  \left[\begin{array}{c|cc}L_{11} & L_{12} & L_{13} \\\hline L_{21} & L_{22}-\lambda & L_{23} \\
 L_{31} & L_{32} & L_{33}-\lambda\end{array}\right]
\\
&& P_3 =\det\left[\begin{array}{c|cc|c}L_{11} & L_{12} & L_{13} &
    L_{14} \\\hline L_{21} & L_{22} & L_{23} & L_{24} \\ L_{31} &
    L_{32} & L_{33} & L_{34} \\\hline L_{41} & L_{42} & L_{43} &
    L_{44}-\lambda\end{array}\right]
\eea
Let us denote $\lambda_{j,\rho}\ ,\ \ \rho = 1,\dots,\ell_j$ the
Lidskii pseudovalues: then the eigenvalue $0$ of $M_0$ splits into $K$
cyclic multiplets, the $j$-th of  which is expandable as a Puiseux
series in $\epsilon^{\frac 1{n_j}}$ and with the leading coefficients
of these series given by the Lidskii pseudovalues.\\
In the above example:\\
\indent we have a $n_1=4$-tuplet of cyclic eigenvalues $\lambda_1(\epsilon) = \lambda_{1,1}\epsilon^{1/4}+\dots.$;\\
\indent we have $\ell_2=2$ $n_2=n_3=2$-tuplets (doublets) with expansion
$\lambda_{2,1}(\epsilon) = \lambda_{2,1}\sqrt{\epsilon}  +\dots $ and
$\lambda_{2,2}(\epsilon) =\lambda_{2,2}\sqrt{\epsilon}+\dots$;\\
\indent we have $\ell_3=1$ eigenvalue with the form $\lambda_{3,1} = \lambda_{3,1}\epsilon + \mathcal O(\epsilon^2)\ .$

It should be clear now that the genericity condition is such to ensure
that the (germ of) the spectral curve at $\epsilon=0$ can be minimally
resolved.

For later purposes we now investigate a bit closer the structure of
the perturbative expansion in the case of a single Jordan block with
nonvanishing Lidskii coefficient.
\bp\label{propspectral}
Let $M(\epsilon) = \mathcal N + \sum_{j=1}^\infty\epsilon^j M_j$ be a
(formal) perturbation of the single nilpotent Jordan  block $\mathcal
N$ of size $n\times n$. Suppose that the Lidskii pseudovalue is
nonzero, namely ${\lambda_1}^n:=(M_1)_{n1}\neq 0$. Then:
\begin{enumerate}
\item There exists a similarity transformation constant in $\epsilon$  which transforms the
  problem in the following perturbation problem
\be
\tilde M(\epsilon) = \le[
\begin{array}{cccccc}
0& \lambda_1 & \lambda_2 &\dots & \lambda_{n-2} & \lambda_{n-1}\\
0&0&\lambda_1 & \ddots & \ddots &  \lambda_{n-2}\\
0&0&0&\ddots& \ddots& \vdots\\
 & \ddots & & & &\lambda_2 \\
& & & & & \lambda_1 \\
0&0&\dots & 0 &0 &0
\end{array}
\ri] +  \epsilon \le[
\begin{array}{cccccc}
\star &\star & \dots & & &\star\\
& & & & &\\
& & & & &\\
& & & & &\\
\star & & \dots & & \\
\lambda_1 & \star & \dots & & &
\end{array}
\ri] + \mathcal O(\epsilon^2)
\ee
 Such similarity is unique up to the centralizer of $\mathcal N$.
\item The coefficients $\lambda_j$, $j=1,\dots n-1$ are the first
  coefficients in the Puiseux expansion of one of the $n$ cyclically
  permuted eigenvalues in powers of $\xi=\epsilon^{\frac 1
  n}$. Furthermore they depend only on the coefficients of $M_1$,
  rationally in $\lambda_1$ and polynomially in the other coefficients.
\end{enumerate}
\ep
{\bf Proof.}\\
Let us define
\bea
&& G:= {\rm diag} (0,1,2,\dots,n-1)\\
&& \xi:= \epsilon^{\frac 1 n}\\
&& \lambda_1:= {(M_1)_{n1}}^{\frac 1 n}.
\eea
We first ``shear'' the perturbation problem
\bea
&& S = (\lambda_1 \xi)^{-G} M (\lambda_1 \xi)^{G} = \nonumber\\
&&=\xi\lambda_1 \mathcal C + \xi^2\le[
\begin{array}{ccccc}
&&&  &\\
& & & &\\
&&&&\\
\star & & &&\\
&\star & & &
\end{array}
\ri]
+ \xi^3 \le[
\begin{array}{ccccc}
&& &  &\\
& & & &\\
\star&&&&\\
 &\star & &&\\
& & \star& &
\end{array}
\ri]
+\dots
+ \xi^n \le[
\begin{array}{ccccc}
\star && &  &\\
& \star& & &\\
&&\ddots&&\\
 & & &\star&\\
& & & &\star
\end{array}
\ri] +  \mathcal O(\xi^{n+1})\nonumber
\eea
where only the marked entries in the expansion above can be nonzero
and the matrix $C$ is the cyclic permutation
\be
\mathcal C:=\left[\begin{array}{ccccc} & 1 &  & &  \\ &
  & 1 &  &  \\ &  & & \ddots  &  \\ &  &  &  & 1 \\ 1 &
  &  & & \end{array}\right]
\ee
Note that the sheared perturbation problem has the following
periodicity and structure
\bea
&&S(\xi) =\sum_{j=1}^\infty \xi^j s_j\mathcal C^j\\
&&s_1:=\lambda_1 \1,\ s_j=\hbox{ diagonal matrices}\ ,j\geq 2\\
&&S(\omega \xi) = \Omega^{-1} S(\xi)\Omega\\
&&\Omega:=\omega^G,\ \ \omega:= {\rm e}^{2i\pi/n}
\eea
Moreover the first $s_2,\dots,s_{n}$ are the stars of the expansion
above and involve only the corresponding entries of $M_1$. In
particular the first $n-j$ coefficients of $s_j$ are zeroes (for
$j\leq n$).\par
The reason for the shearing is that the leading
coefficient of this new perturbation problem has nondegenerate
spectrum $\lambda_1\omega^j,\ j=0,\dots n$ and hence can be dealt with
usual techniques. In fact, let us  introduce the following eigenvector matrix $W$,  of
$\mathcal C$
\be
\mathcal C W = W\Omega^{-1}\ , \ \ W_{ij} = \omega^{-(i-1)(j-1)}\ .
\ee
We claim that we can find a unique perturbative eigenvalue matrix
of the following form
\bea
P(\xi):=\sum_{j=0}^\infty \xi^j \mathcal C^j p_j W\\
p_0=\1\ , p_j=\hbox{ diagonal traceless},
\eea
which diagonalizes $S(\xi)$ with cyclically permuting  eigenvalues
\bea
&&y_j(\xi) =y(\omega ^j \xi)\\
&& y(\xi):=\sum_{j=1}^{\infty} \xi^j\lambda_j\\
&& y_{j+1}(\xi) =y_j(\omega \xi)\ .
\eea
In order to show this we have to solve the following formal power
series identity
\be
S(\xi) P(\xi) = P(\xi)\Lambda(\xi) \ ,\qquad \Lambda(\xi) := y(\Omega^{-1} \xi)\ ,
\ee
where the unknowns are the diagonal traceless matrices $p_j,j\geq 2$
and the scalars $\lambda_j\ ,\ j\geq 2$. Plugging the Ansatz and comparing the
coefficient of the power  $\xi^{K+1}$ we obtain
the recurrence relation
\be
\sum_{j=1}^K s_j\mathcal C^K p_{K-j}W =\sum_{j=1}^{K} \lambda_j
\mathcal C^{K- j} p_{K-j} W\Omega^{-j}\ .
\ee
Solving for $p_{K-1}$ and using $W\Omega^{-j} = \mathcal C^jW$ we
obtain
\be
\lambda_1\bigg( p_{K-1} -\mathcal C^{-1}p_{K-1} \mathcal C\bigg) =
\sum_{j=2}^K \bigg(
\lambda_j \mathcal C^{-j}p_{K-j}\mathcal C^j -\mathcal
C^{-K}s_j\mathcal C^{K} p_{K-j}
\bigg)\ .
\ee
This recurrence relation admits a {\em unique} solution: indeed both sides are
diagonal matrices and we must guarantee that the RHS is traceless
(since the LHS is, whether $p_{K-1}$ is traceless or not). This
condition fixes $\lambda_K$ uniquely to be (recalling that  $p_0=\1$
and all the other $p$'s are traceless)
\be
n\,\lambda_K =  \tr(s_K) + \sum_{j=2}^{K-2} \tr\bigg( \mathcal
C^{-K}s_j\mathcal C^K p_{K-j}\bigg)
\ee
Moreover this shows that $\lambda_1,\dots,\lambda_{n}$ depend only on
$s_1,\dots s_n$ and hence only on the entries of $M_1$ in a polynomial
way w.r.t all entries except $(M_1)_{n1}$, w.r.t which they depend
rationally.\par
We now revert to the original problem by inverting the shearing
\bea
&& M(\epsilon)  = (\lambda_1 \xi)^G S(\xi)(\lambda_1 \xi)^{-G} \ .
\eea
The eigenvector/eigenvalue problem for $S(\xi)$ is conjugated as well
into the following equation which involves only integer powers of $\epsilon$
\be
M(\epsilon) \tilde P(\epsilon) = \tilde P(\epsilon) D(\epsilon)
\ee
where
\bea
\tilde P(\epsilon) &:=& (\lambda_1 \xi)^G  P(\xi) W^{-1} (\lambda_1 \xi)^{-G}
=\sum_{j=0}^\infty \frac 1{{\lambda_1}^j} \mathcal F^j(\epsilon) p_j\\
D(\epsilon) &:=&  (\lambda_1 \xi)^G W\Lambda(\xi) W^{-1}(\lambda_1 \xi)^{-G}
= \sum_{j=1}^{\infty} \frac {\lambda_{j}}{(\lambda_1)^j} \mathcal F^j(\epsilon)
\ .
\eea
Here we have set
\be
\mathcal F(\epsilon):=\lambda_1 \xi \,(\lambda_1 \xi)^G \mathcal
C(\lambda_1 \xi)^{-G}  =
\le[
\begin{array}{ccccc}
&1&&&\\
&&1&&\\
&&&\ddots&\\
(\lambda_1)^n\epsilon&&&&
\end{array}
\ri]
 =
\le[
\begin{array}{ccccc}
&1&&&\\
&&1&&\\
&&&\ddots&\\
(M_1)_{n1}\epsilon&&&&
\end{array}
\ri]
\ee
Note now that the matrix $\tilde P$ is nonsingular and (formally)
analytic in $\epsilon$; the leading coefficient is an invertible
constant upper triangular matrix of the form
\be
\tilde P(\epsilon) = \1 + \sum_{j=1}^{n-1} \mathcal N^j \frac
       {p_j}{(\lambda_1)^j} + \mathcal O(\epsilon) = T+\mathcal O(\epsilon).
\ee
The similarity we are looking for is $(\lambda_1)^{-G}T (\lambda_1)^{G}$, where $T$ is defined here above.
In fact we have the leading coefficient
\bea
 (\lambda_1)^{G}\,(T^{-1} M(\epsilon) T)_0\, (\lambda_1)^{-G} =
(\lambda_1)^{G}\,  ({\tilde P}^{-1} M \tilde P)_0\,
(\lambda_1)^{-G} =(\lambda_1)^{G}\,  D_0\, (\lambda_1)^{-G} =  \\
=(\lambda_1)^{G}\,\le[
\begin{array}{cccccc}
0&1&\frac {\lambda_{2}}{\lambda_1^2} & \frac {\lambda_{3}}{\lambda_1^3} & \dots &
  \frac{\lambda{n-1}}{\lambda_1^{n-1}}\\
&0&1&\frac {\lambda_{2}}{\lambda_1^2}&\ddots& \\
&&&\ddots &\ddots & \frac {\lambda_{3}}{\lambda_1^3}  \\
&&&0&1 & \frac {\lambda_{2}} {\lambda_1^2}\\
&&&&\phantom{\frac {\lambda_{2}}{\lambda_1^2}}&\hspace{-4pt}1\\
&&&&&0
\end{array}
\ri]\, (\lambda_1)^{-G} =  \le[
\begin{array}{cccccc}
0& \lambda_1 & \lambda_2 &\dots & \lambda_{n-2} & \lambda_{n-1}\\
0&0&\lambda_1 & \ddots & \ddots &  \lambda_{n-2}\\
0&0&0&\ddots& \ddots& \vdots\\
 & \ddots & & & &\lambda_2 \\
& & & & & \lambda_1 \\
0&0&\dots & 0 &0 &0
\end{array}
\ri]\ .
\eea
As for the subleading coefficient it is easy to see that the $(n,1)$
entry is transformed to $\lambda_1$, thus completing the proof. Q.E.D.\par\vskip 5pt
Before proceeding we observe that once the perturbation matrix is in
the form advocated in Prop. \ref{propspectral} the
``pseudo-eigenvalue'' problem can be recast as in the following
corollary
\bc
Suppose we have a formal perturbation problem in the form guaranteed
by Prop. \ref{propspectral}; then there exists a
pseudo-eigenvector matrix of the form
\be
P(\epsilon) = \1 + \mathcal O(\epsilon)
\ee
such that
\be
P^{-1}\, M\, P= \sum_{j=1}^\infty \lambda_j \le[
\begin{array}{cccc}
&1&&\\
&&\ddots&\\
&&&1\\
\epsilon&&&
\end{array}\ri]^j
\ee
\ec

\section{Formal Asymptotics}
The cornerstone of our analysis of resonant irregular singularities in
our class is the following theorem, which displays the normal formal
asymptotic form of a kernel solution $\Psi' = A\Psi$ in a sectorial
neighborhood of the singularity.
\bt[Main Theorem]\label{main}
Let $A(x)$ be a $M\times M$ matrix with coefficients polynomial in $x$
of degree at most $r-1$ ($r\geq 1$), and formal series in $x^{-1}$
\be\label{eq:Ax}
A(x) = \sum_{j\leq r} A_j x^{j-1}.
\ee
We assume that the leading coefficient is in Jordan canonical form with the following elementary block structure
\bea
A_r = ({\lambda_1}^{n_1})\cdots({\lambda_s}^{n_{s}})\ ,
\label{eq:Ar}
\eea
where the eigenvalues are not assumed to be distinct.
Under the genericity assumption of Def. \ref{de:generic} on the Lidskii pseudovalues there exists a formal gauge $Y$ analytically invertible at $\infty$ of the form
\be
Y(x) = Y_0 + \sum_{j=1}^{\infty} x^{-j} Y_j\ ,\ \ \det(Y_0)\neq 0\ ,
\ee
such that the gauge transformed connection
\be
D(x):= Y^{-1}A Y - Y^{-1}Y' = {\rm diag}(D_1,\dots,D_s)
\ee
is in block diagonal form according to the minimal  block
decomposition of $A_r$ and where a block of size $n_j$ corresponding to
an eigenvalue $\lambda_j$ has the form
\bea
D_{j} &=&\lambda_j  x^{r-1} +
\frac 1 {n_j x}\sum_{J=0}^{rn_j-1} t_{J,j} {\mathcal H_j(x)}^J - \frac {G_j} x \\
&=& \frac 1
      {n_jx}\sum_{J=0}^{rn_j} t_{J,j}{ \mathcal H_j(x)}^J - \frac {G_j} x \ ,\ \ \
      t_{rn_j}:= n_j\lambda_j\\
&& G_j:={\rm diag}(0,1,\dots,n_j-1)\\
&&\mathcal H_j (x)= \le[
\begin{array}{ccccc}0 & 0 & 0 & 0 & x \\1 & 0 & 0 & 0 & 0 \\0 & 1 & 0
  & 0 & 0 \\0 & 0 & \ddots & 0 & 0 \\0 & 0 & 0 & 1 & 0\end{array}
 \ri]\in \ {\rm Mat}(n_j,n_j,\C)\ .
\eea
The formal gauge $Y$ is uniquely determined. \\
The formal solution of the system can then be written  in the form 
\bea
&&\Psi_{form} = Y(x) \cdot \Psi^{bare}(x) \\
&&\Psi^{bare} = {\rm diag}\le( \exp\le(\sum_{J=1}^{rn_1} \frac {t_{J,1}}J\mathcal
  H^J_1\ri)x^{\frac {t_{0,1}-G_1}{n_1}},\dots,\exp\le(\sum_{J=1}^{rn_s} \frac {t_{J,s}}J\mathcal
  H^J_s\ri)x^{\frac {t_{0,s}-G_s}{n_s}} \ri)\ .
\eea 
\et
The proof of this theorem is accomplished in various steps contained in Propositions \ref{pro:decompose} and \ref{pro:decomp1}, which consist of a refinement of the standard splitting lemma, followed by Prop. \ref{lemma2} together with Prop. \ref{lemmabare}, which specify the canonical form of the connection and its solution.
\br
We are here dealing only with singularities at $x=\infty$ but an
analogous statement can be made for (formal) series in any local
parameter $(x-\gamma)$ by a linear fractional transformation of the variable $x$ in Thm. \ref{main}.
\er
In order to prove Thm. \ref{main} we show that by a  formal
gauge transformation  we can split the connection $A$ into block
diagonal form according to the block diagonal structure of the Jordan
canonical form of its leading coefficient (under the assumed
genericity assumption for the subleading coefficient).

As a preliminary step we apply the standard splitting lemma
\cite{wasow} which guarantees the existence of a formal gauge  $Y$
such that the transformed connection
\begin{eqnarray*}
B^{(1)}(x)= Y^{-1}A Y - Y^{-1}Y'\ ,
\end{eqnarray*}
 is block diagonal and each block has a leading term with only one eigenvalue
\begin{eqnarray*}
B^{(1)}=&\pmatrix{B_1^{(1)}&0&\ldots &0\cr
           0&B_2^{(1)}&\ldots &0\cr
           \vdots&\vdots&\ddots&\vdots\cr
           0&\ldots &0&B_K^{(1)}\cr}\
\end{eqnarray*}
 $K$ is the number of groups of coinciding eigenvalues in (\ref{eq:Ar}), the size of each block $B_i^{(1)}$ is the rank of the nilspace of the matrix $A_r$ with the same eigenvalue and different block correspond to distinct eigenvalues of $A_r$.

This can be shown by using the following well-known result
inductively.
\begin{proposition}Suppose that the eigenvalues of $A_r$ in the series
of the form (\ref{eq:Ax}) has two groups of eigenvalues
$\lambda_1,\ldots,\lambda_p$ and $\lambda_{p+1},\ldots,\lambda_K$
such that $\lambda_i\neq\lambda_j$ for $i\leq p$ and $j>p$. Then
$A_r$ is similar to a block diagonal matrix
\begin{eqnarray*}
A_r=\pmatrix{A_1&0\cr
           0&A_2\cr}\
\end{eqnarray*}
where $A_1$, $A_2$ are of dimensions $p \times p$ and
$(k-p)\times (k-p)$ respectively.

Furthermore  there exists a formal power series $Y$ such that
\begin{eqnarray*}
B^{(1)}(x)= Y^{-1}A Y - Y^{-1} Y'\ ,
\end{eqnarray*}
has the block diagonal form
\begin{eqnarray*}
B^{(1)}=\pmatrix{B_1^{(1)}&0\cr
           0&B_2^{(1)}\cr}\
\end{eqnarray*}
where $B_1^{(1)}$, $B_2^{(1)}$ are square matrices of dimensions
$m_1\times m_1$ and $m_2\times m_2$ respectively.
\end{proposition}
The proof of this proposition can be found for example in \cite{wasow} or \cite{sibuya}. \par\vskip 5pt

In view of the above splitting lemma we can restrict ourselves to the
case where the Jordan form of $A_r$ has only one eigenvalue, that is
\begin{eqnarray}\label{eq:Ar1}
A_r&=&(\lambda^{n_1})\cdots (\lambda^{n_s})
\end{eqnarray}
 and, without lost of generality, we can assume that $\lambda=0$ by performing a scalar gauge transformation ${\rm e}^{-\frac {\lambda}{r+1} x^{r+1}}\1$. We
shall now prove the following proposition that shows decomposability
of  $A(x)$ in block diagonal form under the genericity assumption in
Def.
\ref{de:generic}.

\bp
\label{pro:decompose} Suppose $A_r$ in the
 series (\ref{eq:Ax}) is in Jordan canonical form
\begin{eqnarray*}
A_r&=&(0^{n_1})\cdots(0^{n_s})
\end{eqnarray*}
If the Lidskii pseudovalues of $A_{r-1}$ are generic in the sense
of Def. \ref{de:generic}, then there exists a formal power
series
\begin{eqnarray*}
Y(x) = Y_0 + \sum_{j=1}^{\infty} x^{-j} Y_j\ ,\ \ \det(Y_0)\neq 0\ ,
\end{eqnarray*}
such that
\begin{eqnarray*}
B(x)= Y^{-1}A Y - Y^{-1}Y'\ ,
\end{eqnarray*}
is a formal power series in block diagonal form according to the
block decomposition of $A_r$.
\ep

The above is a consequence of the following
\begin{proposition}\label{pro:decomp1}
If the Lidskii pseudovalues of $A_{r-1}$ are generic, then we can
find a gauge $H$ linear in $x^{-1}$ such that the first two
leading terms of
\begin{eqnarray}\label{eq:tildeA}
\tilde{A}(x)= H^{-1}A H - H^{-1}H'\ ,
\end{eqnarray}
have the following form
\begin{eqnarray}\label{eq:tildeAr}
\tilde{A}_r&=&(0^{n_1})\cdots(0^{n_s}) =A_r\\
\tilde{A}_{r-1}&=&\pmatrix{\tilde{A}_{r-1,1}&0&\ldots &0\cr
           0&\tilde{A}_{r-1,2}&\ldots &0\cr
           \vdots&\vdots&\ddots&\vdots\cr
           0&\ldots &0&\tilde{A}_{r-1,s}\cr}\
\end{eqnarray}
where $\tilde{A}_{r-1,i}$ are square matrices of dimensions
$n_i\times n_i$ such that the only non-zero entries of
$\tilde{A}_{r-1,i}$ are the
$\left(\tilde{A}_{r-1,i}\right)_{n_i,1}$, that is, the bottom left
hand corner.
\end{proposition}
Proof. Let $H=H_0+H_1x^{-1}$. We first partition $H_0$ and $H_1$
into $s\times s$ blocks according to the decomposition of $A_r$.
We will denote these blocks by $H_{jk}^{(0)}$ and $H_{jk}^{(1)}$
respectively. The block $H_{jk}^{(i)}$ is a  rectangular matrix of
dimension $n_j\times n_k$. We shall similarly partition $A_{r-1}$
into $s\times s$ blocks and call these $A_{jk}$ for simplicity (i.e. we suppress the index $r-1$).

The coefficients of $x^r$ and $x^{r-1}$ in (\ref{eq:tildeA}) can
now be written in the form
\begin{eqnarray*}
A_rH_0-H_0A_r&=&0 \\
A_rH_1-H_1A_r&=&A_{r-1}H_0-H_0\tilde{A}_{r-1}
\end{eqnarray*}
We can write these equations in block form, in which the $jk^{th}$
block of both sides are given by
\begin{eqnarray}\label{eq:recurH}
\mathcal N_jH_{jk}^{(0)}-H_{jk}^{(0)}\mathcal N_k&=&0 \nonumber\\
\mathcal N_jH_{jk}^{(1)}-H_{jk}^{(1)}\mathcal
N_k&=&\sum_{l=1}^sA_{jl}H_{lk}^{(0)}-H_{jk}^{(0)}\tilde{A}_{r-1,k}
\end{eqnarray}
where $\mathcal N_j$ is the $n_j\times n_j $ dimensional shift matrix
\begin{eqnarray*}
\mathcal N_j:= \left[\begin{array}{cccc}0 & 1 & 0 & 0 \\0 & 0 &
\ddots & 0 \\0 & 0 & 0 & 1 \\0 & 0 & 0 & 0\end{array}\right]\ .
\end{eqnarray*}

We want  to find $H_0$ and $H_1$ such that (\ref{eq:recurH})
is satisfied with $\tilde{A}_{r-1}$ given by (\ref{eq:tildeAr}).
Let us denote the following linear operator from the space of
$n_j\times n_k$ rectangular matrix to the same space by $\mathcal
L_{jk}$
\begin{eqnarray}\label{eq:L}
\mathcal L_{jk}:{\bf Mat}(n_j,n_k)\rightarrow {\bf Mat}(n_j,n_k)\\
 X\mapsto\mathcal N_jX-X\mathcal N_k
\end{eqnarray}
The first equation in (\ref{eq:recurH}) means that $H_{jk}^{(0)}$
is in the kernel of $\mathcal L_{jk}$, while right hand side of
the second equation in (\ref{eq:recurH}) has to be in the image of
$\mathcal L_{jk}$. To prove the proposition, we will need to know the
structure of the kernel and image of $\mathcal L_{jk}$ and show that under
the genericity assumption in Def.  \ref{de:generic}, (\ref{eq:recurH}) is
solvable.

The following can be verified by
straightforward calculation
\begin{lemma}\label{le:ker}
The kernel of $\mathcal L_{jk}$ is of dimension $min(n_j,n_k)$ and
it is spanned by $n_j\times n_k$ matrices $J$ of the form
\begin{eqnarray}\label{eq:ker}
J&=&\pmatrix{0&0&\ldots &J_1&\ldots&J_{n_j-1} &J_{n_j}\cr
           0&0&\ldots &0&J_1&\ldots &J_{n_j-1}\cr
           \vdots&\vdots&\ddots&\vdots&\vdots&\vdots&\vdots\cr
           0&\ldots &0&0&0&0&J_1\cr}\ \quad n_j<n_k \\
J&=&\pmatrix{J_1&\ldots&J_{n_k-1} &J_{n_k}\cr
           0&J_1&\ldots &J_{n_k-1}\cr
           \vdots&\vdots&\ddots&\vdots\cr
           0&\ldots &0&J_1\cr
           \vdots&\vdots&\ddots&\vdots\cr
           0&\ldots &0&0\cr}\ \quad n_k<n_j
\end{eqnarray}
where $J_i$ are arbitrary constants. That is, the kernel is
spanned by $n_j\times n_k$ matrices with top right hand corner of
the form
\begin{eqnarray*}
\tilde{J}&=&\pmatrix{J_1&\ldots&J_{m-1} &J_{m}\cr
           0&J_1&\ldots &J_{m-1}\cr
           \vdots&\vdots&\ddots&\vdots\cr
           0&\ldots &0&J_1\cr}\
\end{eqnarray*}
and all other entries zero, where $m=min(n_j,n_k)$
\end{lemma}
The next lemma (left as exercise) characterizes the image of the operator $\mathcal L_{jk}$.
\begin{lemma}\label{le:image}
The image of $\mathcal L_{jk}$ is of codimension $min(n_j,n_k)$ and it is
given by $n_j\times n_k$ matrices $Q$ such
that
\begin{eqnarray}\label{eq:image}
\sum_{i=1}^p Q_{n_j-p+i,i}=0,\quad p=1,\ldots ,m
\end{eqnarray}
where $m=min(n_j,n_k)$.
\end{lemma}
This means that the ``dual diagonals'' to the ones appearing in the
characterization of the kernel of $\mathcal L_{jk}$ are
``traceless''.\\
We now look at the term  $V^{jk}=\sum_{l=1}^sA_{jl}H_{lk}^{(0)}$
in (\ref{eq:recurH}). To simplify the notation, we shall denote
the entry in the bottom left  corner of $A_{jl}$ by $a(j,l)$ and
we write $H_{lk}^{(0)}$ in the form (\ref{eq:ker})
denoting the elements on the diagonals by $h_i^{jk}$ and
$h_i^{jk}=0$ if $i>min(n_j,n_k)$ or $i\leq 0$. (i.e., the $J_i$ in
(\ref{eq:ker})). Let $m_{jk}=min(n_j,n_k)$ and consider the terms
\begin{eqnarray*}
\sum_{i=1}^p V_{n_j-p+i,i}^{jk},\quad p=1,\ldots ,m_{jk}
\end{eqnarray*}
in $V^{jk}$, as these terms determine whether the right hand side of (\ref{eq:recurH}) is in the
image of $\mathcal L_{jk}$.

Let $c_{jk}=min(n_j-n_k,0)$, a simple calculation then shows that
\begin{eqnarray}\label{eq:V}
\sum_{i=1}^p
V_{n_j-p+i,i}^{jk}=\sum_{l=1}^sa(j,l)h_{c_{jk}+p}^{lk}+R_p^{k},\quad
p=1,\ldots ,m_{jk}
\end{eqnarray}
where $R_p^{k}$ is a term that involves only $h_q^{lk}$ for $l=1,\ldots,s$ and $q<p+c_{jk}$ and $h_q^{lk}=0$ if
$q\leq 0$.
Recall that
$l_1$ is the number of blocks with the same size $n_1$, $l_2$ is the number of blocks with the same size
$n_{l_2}$, etc. Let $\tilde{H}^k$ be the column vector with entries
\begin{eqnarray*}
\left(\tilde{H}^k\right)_{\sum_{j=1}^i(s-w_{j-1})+q}=h_i^{qk}, \quad
i=1,\ldots, n_k,\quad q=1,\ldots,w_{i-1}
\end{eqnarray*}
where $w_i=\sum_{q=1}^{p_i} l_q$ and $p_i$ is the biggest number
such that $n_{w_i}>i$.

More explicitly, $\tilde{H}^k$ is the column vector
\begin{eqnarray}\label{eq:tildeH}
\tilde{H}^k=\pmatrix{h_1^{1,k}\cr
                   h_1^{2,k}\cr
                   \vdots\cr
                   h_1^{s,k}\cr
                     \hline \cr
                     h_i^{1,k}\cr
                     \vdots\cr
                     h_i^{w_{i-1},k}\cr
                     \hline \vdots\cr
                     \hline
                      h_{n_k}^{1,k}\cr
                      \vdots\cr
                       h_{n_k}^{w_{n_{k-1}},k}\cr}\
\end{eqnarray}
Now let $\tilde{V}^k$ be the following column vector
\begin{eqnarray*}
\left(\tilde{V}^k\right)_{\sum_{j=1}^i(s-w_{j-1})+q}=\sum_{m=1}^i
V_{n_q-i+m,m}^{qk},\quad i=1,\ldots ,n_k ,\quad q=1,\ldots,w_i
\end{eqnarray*}
That is, $\tilde{V}^k$ is obtained from the vector $\tilde{H}^k$ by replacing $h_i^{qk}$ by the corresponding
`sums of diagonal' in $V^{jk}$.
\begin{eqnarray}\label{eq:tildeV}
\tilde{V}^k=\pmatrix{V_{n_{1},1}^{1k}\cr
                   V_{n_{2},1}^{2k}\cr
                   \vdots\cr
                   V_{n_{s},1}^{sk}\cr
                   \hline  \vdots\cr
                   \hline
                     \sum_{q=1}^i V_{n_{1}-i+q,q}^{1k}\cr
                     \vdots\cr
                     \sum_{q=1}^i V_{n_{w_i}-i+q,q}^{w_{i-1},k}\cr
                     \hline \vdots\cr
             \hline
                      \sum_{q=1}^{n_k} V_{n_1-n_k+q,q}^{1k}\cr
                      \vdots\cr
                       \sum_{q=1}^{n_k} V_{q,q}^{w_{n_{k-1}},k}\cr}\
\end{eqnarray}
If we denote the bottom left hand diagonal by the 'first
diagonal', the next one by the 'second diagonal' and so forth, and
call the sum of their elements the 'trace of the diagonal', then
the $i^{th}$ block in $\tilde{V}^k$ consists of the traces of the
$i^{th}$ diagonals.

We can now write the equations (\ref{eq:V}) for $j=1,\ldots, s$ in
matrix form
\begin{eqnarray}\label{eq:matrixV}
U^k\tilde{H}^k=\tilde{V}^k
\end{eqnarray}
From (\ref{eq:V}), we see that the matrix $U^k$ is of the form
\begin{eqnarray*}
U^k=\pmatrix{U_{1}^k&0&\ldots &0 \cr
           \star&U_{2}^k&\ldots &0\cr
           \vdots&\vdots&\ddots&\vdots\cr
           \star&\ldots &\star&U_{n_k}^k\cr}\
\end{eqnarray*}
where $\star$ denotes some non-zero entries. The matrices $U_i^k$
are given by
\begin{eqnarray*}
U_i^k=\left[\begin{array}{ccc|cc}a(1,1)&\ldots
&a(1,w_{n_k-1})&0&\\
           \vdots&\ddots&\vdots&\vdots&\\
           a(w_{n_k-1},1)&\ldots&a(w_{n_k-1},w_{n_k-1})&0&\\
           \hline 0&\ldots &0&0\end{array}\right]
\end{eqnarray*}
that is, $U_i^k$ is the $w_{i-1}^2$ matrix that has the principal
block submatrix of the Lidskii matrix of size $w_{n_k-1}^2$ on its
top left hand corner and zero elsewhere.

We now consider the term $H_{jk}^{(0)}\tilde{A}_{r-1,k}$ in
(\ref{eq:recurH}). Let $\lambda_i$ be the bottom left hand entry
of $\tilde{A}_{r-1,i}$, that is, its only non-zero entry. We can
now write this term in matrix multiplication form as in
(\ref{eq:matrixV})
\begin{eqnarray}\label{eq:matrixV2}
\tilde{U}^k\tilde{H}^k=\tilde{V}^k
\end{eqnarray}
where $\tilde{U}^k$ is the following matrix
\begin{eqnarray*}
\tilde{U}^k=\pmatrix{\tilde{U}_{1}^k&0&\ldots &0\cr
           \star&\tilde{U}_{2}^k&\ldots &0\cr
           \vdots&\vdots&\ddots&\vdots\cr
           \star&\ldots &\star&\tilde{U}_{n_k}^k\cr}\
\end{eqnarray*}
where $\tilde{U}_i^k$ is the diagonal matrix given by
\begin{eqnarray*}
\left(\tilde{U}_i^k\right)_{jj}&=&0, j\leq w_{n_k+1} \\
\left(\tilde{U}_i^k\right)_{jj}&=&\lambda_k, j> w_{n_k+1}
\end{eqnarray*}
The right hand side of (\ref{eq:recurH}) can then be written in
the matrix multiplication form
\begin{eqnarray*}
\left(U^k-\tilde{U}^k\right)\tilde{H}^k=\tilde{V}^k, \quad
k=1,\ldots, s
\end{eqnarray*}
By lemma \ref{eq:image}, the condition that the right hand side of
(\ref{eq:recurH}) is in the image of the operator $\mathcal
L_{jk}$ means that the determinant of $U^k-\tilde{U}^k$ has to be
zero. This determinant is given by
\begin{eqnarray*}
\det\left(U^k-\tilde{U}^k\right)=\prod_{i=1}^{n_k}\det\left(U_i^k-\tilde{U}_i^k\right)
\end{eqnarray*}
The first non-zero entry of $\tilde{U}_i^k$ is on the
$w_{n_k-1}+1$ row while the first zero entry of $U_i^k$ is on the
$w_{n_k}$ row, therefore, $U_i^k-\tilde{U}_i^k$ is given by
\begin{eqnarray*}
U_i^k-\tilde{U}_i^k=\left[\begin{array}{cc|ccc|c}
a(11)&\ldots
&a(1,w_{n_k}+1)&\ldots&a(1,w_{n_k-1})&0\\
           \vdots&\ddots&\vdots&\ddots&\vdots&\vdots\\
          \hline a(w_{n_k}+1,1)&\ldots&a(w_{n_k}+1,w_{n_k}+1)-\lambda_k&\ldots&a(w_{n_k}+1,w_{n_k-1})&0\\
          \ldots&\ddots&\ldots&\ddots&\ldots&\vdots\\
          a(w_{n_k-1},1)&\ldots&a(w_{n_k-1},w_{n_k}+1)&\ldots&a(w_{n_k-1},w_{n_k-1})-\lambda_k&0\\
          \hline 0&\ldots&0&\ldots&0 &-\lambda_kI\end{array}\right]
\end{eqnarray*}
where $I$ is the identity matrix of dimension
$\left(w_{i-1}-w_{n_k-1}\right)^2$.

We see that the determinants
$\det\left(U_i^k-\tilde{U}_i^k\right)$ are just products of
$\lambda_k$ and the pseudo-characteristic polynomials of the
Lidskii submatrices. Therefore the genericity assumption in
Def. \ref{de:generic} would imply that the determinant of
$U^k-\tilde{U}^k$ is zero and that the right hand side of
(\ref{eq:recurH}) lies in the image of $\mathcal L_{jk}$. Q.E.D.\par\vskip 5pt

We can now prove  Prop.\ref{pro:decompose} as a corollary of
Prop.\ref{pro:decomp1}. We can assume that the genericity
condition holds and that $A_r$ and $A_{r-1}$ are given by
(\ref{eq:tildeAr}) in proposition \ref{pro:decomp1}.

Let $Y$ be a power series such that
\begin{eqnarray*}
Y(x) = Y_0 + \sum_{j=1}^{\infty} x^{-j} Y_j\ ,\ \ \det(Y_0)\neq 0\ ,
\end{eqnarray*}
and
\begin{eqnarray}\label{eq:B}
B(x)= Y^{-1}A Y - Y^{-1}Y'\ ,
\end{eqnarray}
for some Laurent series $B$ in $x$. The coefficient of $x^{r-i}$
in (\ref{eq:B}) then gives
\begin{eqnarray*}
A_rY_i-Y_iA_r=A_{r-1}Y_{i-1}-Y_{i-1}A_{r-1}+\sum_{l=2}^{i}A_{r-l}Y_{i-l}-Y_{i-l}B_{r-l}-
(r-i)Y_{i-r-1}
\end{eqnarray*}
where we have set for convenience $Y_k\equiv 0$ if $k<0$.

Assuming that $B$ is in block diagonal form, we will show that
(\ref{eq:B}) is solvable. We first write these equations into block
form, as in the proof of proposition
\ref{pro:decomp1}, in which the $(j,k)$ blocks of both sides are
given by
\begin{eqnarray}\label{eq:recurY}
\mathcal N_jY_{jk}^{(i)}-Y_{jk}^{(i)}\mathcal
N_k&=&A_{r-1,j}Y_{jk}^{(i-1)}-Y_{jk}^{(i-1)}A_{r-1,k}+\left(\sum_{l=2}^{i}A_{r-l}Y_{i-l}-
(r-i)Y_{i-r-1}\right)_{jk}\nonumber
\\&-&\sum_{l=2}^iY_{jk}^{(i-l)}B_{kk}^{(r-l)}
\end{eqnarray}
where $X_{jk}$ denotes the $jk^{th}$ block of $X$. Provided that
the right hand side is in the image of $\mathcal L_{jk}$,
(\ref{eq:recurY}) determines $Y_{jk}^{(i)}$ up to the addition of
an element in the kernel of $\mathcal L_{jk}$. We now  are going to
show that this arbitrariness in the solution of the $i^{\hbox{th}}$
equation can be used to guarantee the solvability of the $(i+1)^{th}$
equation, namely in such a way that  the right hand side of
(\ref{eq:recurY}) lies in the image of $\mathcal L_{jk}$ in
the $(i+1)^{th}$ equation.

Let $J$ be a matrix  in the kernel of $\mathcal L_{jk}$ as in
(\ref{eq:ker}), and $A_{r-1,j}$ be the $n_j\times n_j$ matrix
\begin{eqnarray*}
A_{r-1,j}=\pmatrix{0&0&\ldots &0\cr
           0&0&\ldots &0\cr
           \vdots&\vdots&\ddots&\vdots\cr
           \lambda_j&\ldots &0&0\cr}\
\end{eqnarray*}
where $\lambda_j\neq 0$. When the genericity condition is
satisfied, we also have $\lambda_j\neq\lambda_k$ if $n_j=n_k$, $j\neq k$.

Recall that the bottom left hand diagonals are called the ``first
diagonal'', the next one  'second diagonal' and so forth, and
the sums of their elements are called the ``traces of the
diagonal''. A simple calculation shows that the trace of the
$l^{th}$ diagonal of
$A_{r-1,j}Y_{jk}^{(i-1)}-Y_{jk}^{(i-1)}A_{r-1,k}$ is
\begin{eqnarray*}
\lambda_jJ_{c(k,j)+l}-\lambda_kJ_{c(j,k)+l}
\end{eqnarray*}
where $c(l,m)=min(n_l-n_m,0)$ and $J_k=0$ if $k\leq 0$.
Note that if the blocks have the same size then $c(l,m)=0$, but then
$\lambda_j\neq \lambda_k$ due to our genericity assumption: if they
have different sizes then either $c(l,m)$ or $c(m,l)$ is strictly
negative, and hence the system for the $J_k$'s is triangular with
determinant a power of $\lambda_j$ (or $\lambda_k$, depending on which
one is the ``long side''). In any case
 we  see
that the traces of the first $m_{jk}$ diagonals can be chosen to
be any value by a suitable choice of $J$, therefore one can always
choose $Y_{jk}^{(i-1)}$ such that the right hand side of
(\ref{eq:recurY}) lies in the image of $\mathcal L_{jk}$. Hence
(\ref{eq:B}) is solvable with a block-diagonal $B$. Q.E.D.\par\vskip 5pt

By application of Prop. \ref{pro:decompose} and
Prop. \ref{pro:decomp1} we can reduce to analyzing a single block of
the final connection and we accomplish this final step in the
following proposition.

\bp\label{lemma2}
Let $A(x)$ be a $n\times n$  matrix of degree $r-1$ and formal Laurent
tail, such that the leading coefficient is $\lambda \1$ plus a nilpotent matrix of rank
$n-1$. Suppose also that the Lidskii coefficient of the subleading
term (in the constant gauge in which $A_r$ is in Jordan canonical
form) does not vanish.
 \be
A(x) = x^{r-1} A_r + x^{r-2} A_{r-1} + \dots
\ee

Then there exists a  nonsingular  formal gauge $Y(x)$ of the form
\bea Y(x) =  C\le(\1 + \mathcal O(x^{-1})\ri)\ , \eea such that
the connection $\pa_x - A(x)$ is formally gauged to the connection
$\pa_x - D(x)$ of the form
\bea
\label{shear} D(x) = \frac
{t_{rn}}n x^{r-1} + \frac 1 {nx}\sum_{j=0}^{rn-1} t_j \mathcal H^j
- \frac G x \\
t_{rn}:= n\lambda\ , \
G:={\rm diag} (0,1,2,\dots,n-1)\nonumber \\
\mathcal H = \le[
\begin{array}{ccccc}0 & 0 & 0 & 0 & x \\1 & 0 & 0 & 0 & 0 \\0 & 1 & 0
  & 0 & 0 \\0 & 0 & \ddots & 0 & 0 \\0 & 0 & 0 & 1 & 0\end{array}
 \ri]\nonumber
\eea
\ep
{\bf Proof}. First of all we assume without loss of generality that
 $\lambda =0$ since we can gauge it away by a scalar gauge
 transformation. Next, we can conjugate the
 connection by a constant gauge so that the leading coefficient
 becomes the nilpotent Jordan block $\mathcal N$. Then, as shown
 in Prop. \ref{propspectral}, we can perform a second conjugation
 (constant in $x$) which recasts the problem in the form
\be
A(x)\mapsto A(x) = x^{r-1}
 \le[
\begin{array}{cccccc}
0& \lambda_1 & \lambda_2 &\dots & \lambda_{n-2} & \lambda_{n-1}\\
0&0&\lambda_1 & \ddots & \ddots &  \lambda_{n-2}\\
0&0&0&\ddots& \ddots& \vdots\\
 & \ddots & & & &\lambda_2 \\
& & & & & \lambda_1 \\
0&0&\dots & 0 &0 &0
\end{array}
\ri] +  x^{r-2}\le[
\begin{array}{cccccc}
\star &\star & \dots & & &\star\\
& & & & &\\
& & & & &\\
& & & & &\\
\star & & \dots & & \\
\lambda_1 & \star & \dots & & &
\end{array}
\ri] + \dots\ ,\label{cano}
\ee
where the coefficients $\lambda_j$ are defined as the coefficients of the formal expansion
of the {\bf eigenvalues} of $A(x)$ in Puiseux series of $q=x^{1/n}$
\bea
&& y_j(q) = y (\omega^j q)\\
&& y(q) =q^{(r-1)n} \sum_{j=1}^\infty \lambda_j q^{-j} = x^{r-1}
\sum_{j=1}^\infty \lambda_j x^{-\frac j n }
\eea
In fact we should identify the matrix $x^{1-r} A(x) $ with the
perturbation problem $M(\epsilon)= M(1/x)$ used in Prop. \ref{propspectral}.

We use the same symbol $A(x)$ for this new connection in ``canonical''
form to economize on notation.
At this point we can perform the shearing transformation as done in
Prop. \ref{propspectral}: the difference is that now the
transformation is a change of gauge and not merely a conjugation.
Therefore we  introduce another formal connection $B$ which will be a
formal Laurent series in $q= x^{1/n}$ (by choosing in an arbitrary but
fixed way the determination of the  root)
\bea
&&\widetilde B(q) := SAS ^{-1} +  S'S^{-1} \\
&&  S:= q ^G\ ,\ \ G:={\rm diag} (0,1,2,\dots,n-1)\ .
\eea
Finally we change the variable of differentiation from $x$ to $q=x^{1/n}$
\be
\tilde \Psi'(q) = n q^{n-1} \widetilde B(q)\tilde \Psi  =: n B(q) \tilde \Psi.
\ee
where the matrix $B(q)$, as a consequence of the shearing, has the following structure and enjoys the following properties.
\bea
 B(q)& =& q^{R-1} \le[\overbrace{\lambda_1 \mathcal C }^{=:B_R}+ \frac 1 q B_{R-1} + \dots \ri]  =
 \sum_{j=-\infty}^{R} B_j q^{j-1} \nonumber\\
   & =& q^{R-1} \lambda_1 \mathcal C + q^{R-2}\le[
\begin{array}{ccccc}
&&\lambda_2 &  &\\
& & &\ddots  &\\
&&&&\lambda_2\\
\star & & &&\\
&\star & & &
\end{array}
\ri] + \dots
+ q^{R-n} \le[
\begin{array}{ccccc}
&& &  &\lambda_{n-1}\\
\star& & & & \\
&\star&&&\\
 & &\ddots &&\\
& & & \star&
\end{array}
\ri]
+\dots \nonumber\\
 && \mathcal C :=\left[\begin{array}{ccccc}0 & 1 & 0 & \dots & 0 \\0 &
 0 & 1 & 0 & 0 \\0 & 0 & 0 & \ddots & 0 \\0 & \dots & 0 & 0 & 1 \\ 1 &
 0 & 0 & 0 & 0\end{array}\right]\\
 && B_{j} = b_j \mathcal C^{j} \ , \hbox{ $b_j$ diagonal matrices}\nonumber\\
 && R=rn-1\nonumber \\
 && (b_{R-j})_{\ell\ell}  = \lambda_j\ , j=1,\dots n-1,\ \ell =1,\dots, n-j \nonumber\\
&&
 \omega B(\omega  q ) = \Omega B(q)\Omega^{-1}\ ,\nonumber\\
 && \Omega:=
 diag(1,\omega,\omega^2,\dots, \omega^{n-1}),\ \omega:={\rm
 e}^{2i\pi/n}  \label{sheared}
 \eea
 Since the leading coefficient is nondegenerate we may apply the standard theory of
 asymptotic expansions as in \cite{wasow,sibuya,JMU} stating that we can find a formal solution in the form
 \bea
\widetilde  \Psi & &Z(q) {\rm e}^{T(q)}\nonumber \\
 Z &=& Z_0 + \mathcal O(q^{-1})\nonumber\\
 T(q) &=& \sum_{j=1}^R \frac 1 j{T_j}q^j + T_0 \ln (q)\nonumber
 \eea
  However it is more suitable to derive a slightly improved statement as in the following
 \bl\label{lemma3}
There exists a (unique) formal  solution satisfying the conditions
\bea
\Psi (q)&=& Z(q){\rm e}^{nT(q)}\\
T(q) &=& \sum_{j=1}^{R} \frac {t_j}j \Omega^j q^j + t_0 \ln (q)\ ,\hbox{ $t_j$ scalars}\\
Z(q)&=&\Omega^{-1}Z(\omega q) \mathcal C\\
Z (\infty)&=&  W \ ,
\eea
where $W$ is the following eigenvector matrix of $C$
\be
\mathcal C W = W\Omega^{-1} \ , \ W_{ij} = \omega^{-(i-1)(j-1)}
\ee
\el
{\bf Proof}.\\
This lemma is the counterpart of Prop. \ref{propspectral} in the
setting of formal gauge equivalence.
We start by remarking that the "periodicity" properties of $Z$ and
 $T'$ are compatible with the periodicity properties of $B$
\bea
nB = n Z T' Z^{-1} +  Z'Z^{-1}\\
\omega T'(\omega q) = \mathcal C T'(q) \mathcal C^{-1}\ .
\eea
The differential equation is conveniently rewritten as
\be
\frac 1 n Z' = BZ- ZT'\ .
\ee
Let us expand $Z$ in (formal) Laurent series
\be
Z(q) = \sum_{j=0}^{\infty} q^{-j} Z_j\ ,\ \
Z_0:= W\ .
\ee
The periodicity of $Z$
 implies the following structure for the coefficients of the power series expansion
\be
Z_j = \mathcal C^j z_j W\ ,\ \ z_j=\hbox{ diagonal matrices.}
\ee
We first claim that any such series can be factorized as follows
\bea
\sum_{j=0}^{\infty}q^{-j} \mathcal C^j z_j W &=&
 \sum_{\ell=0}^{\infty} q^{-\ell} \mathcal C^{\ell} p_\ell W \times
 \sum_{k=0}^{\infty} q^{-k} \gamma_k \Omega^{-k}=\\
&=& \sum_{\ell=0}^{\infty} q^{-\ell} \mathcal C^{\ell} p_\ell W
 \times{\rm exp}\le(\sum_{k=0}^{\infty} q^{-k} \delta_k
 \Omega^{-k}\ri)\\
\tr(p_\ell) &=& 0\ ,\ \ \gamma_0 = 1\ ,\ \ p_0=\1\ ,\ \
\gamma_j=\hbox{ scalars }. \eea Indeed, by comparing the two sides
term-wise in the expansion one finds the recursion relations \bea
n\gamma_j  &=& \tr(z_j)\ ,\\
p_j &=& z_j - \gamma_j\1  - \sum_{k=1}^{j-1} p_k\gamma_{j-k}\ .
\eea Once the coefficients $\gamma_j$ are known the coefficients
$\delta_k$
 in the exponential of last expression can be similarly defined
 in a recursive and unique fashion.\\
If we plug this factorized expression in the equation and compare the
 terms of the same power we obtain for the coefficient of $q^{K-1}$
\bea
\sum_{j=0}^{K} \le( B_{R-j} \mathcal C^{K-j} p_{K-j} W -  t_{R-j} \mathcal C^{K-j}p_{K-j} W \Omega^{-j-1}\ri) =0
\ , \ \ K=0,\dots R.
\eea
We recognize that this recurrence relation (for $K\leq R$) is exactly
the same appearing in the proof of Prop. \ref{propspectral}, which
implies that $t_R=\lambda_1$.
Since $B_R=t_R\mathcal C$ and $\mathcal C^{-j}W = W \Omega^j$ we obtain an equation for $p_K$
\be
p_K - \mathcal C^{-1} p_K \mathcal C = \frac 1{t_R}\sum_{j=1}^K \le(
 t_{R-j} \mathcal C^{-j-1}p_{K-j} \mathcal C^{j+1} - \mathcal C^{-K-1}
 B_{R-j} \mathcal C^{K-j} p_{K-j}\ri) \ .
\ee
This equation fixes the diagonal  matrix $p_K$ up to addition of a
 multiple of the identity (i.e. modulo the trace), but since $p_K$ is
 traceless, this equation suffices in fixing it unambiguously.
 Moreover this equation also determines $t_{R-K}$ from the fact that
 the LHS is by default a traceless matrix. Hence
\be
t_{R-K} =  \frac 1 n \tr\le( \mathcal C^{-K-1}B_{R-K} +
 \sum_{j=1}^{K-1}  \le(  \mathcal C^{-K-1} B_{R-j} \mathcal C^{K-j}
 p_{K-j} \ri)\ri)\ .
\ee
This also shows that $t_{R-K}= \lambda_{K+1}$ since the above is the
same recurrence relation that defines $\lambda_K$, $K\leq R$.

For the coefficients of $q^{-K-1}$, $K\geq 1$ we have instead the
following relations (we set $t_{-j}\equiv 0$ for convenience in
writing the following formula)
\bea
\sum_{j=-K}^{R} \bigg( t_j \mathcal C^{K+j}p_{K+j}\mathcal C^{-j} W- B_j \mathcal C^{K+j} p_{K+j} W \bigg)= \\
= \frac K n  \mathcal C^{K} p_{K}W  +
 \frac  1 n \sum_{\ell=0}^{K-1} (\ell-K) \delta_{\ell-K} \mathcal C^{\ell} p_{\ell}\mathcal C^{K-\ell} W
\eea
Multiplying both sides on the left by $\mathcal C^{-K}$ and on the right by $W^{-1}$ we obtain
\bea
t_R p_{K+R}- t_R \mathcal C^{-1} p_{K+R} \mathcal C &=&- \frac K n  p_{K} +   \sum_{j=-K}^{R-1}\bigg(
  t_j \mathcal C^j p_{K+j} \mathcal C^{-j} - \mathcal C^{-K} B_j \mathcal C^{K+j} p_{K+j}
  \bigg)+ \\
  &&-
\frac 1 n  \sum_{\ell=0}^{K-1} (\ell-K) \delta_{K-\ell} \mathcal C^{\ell} p_{\ell}\mathcal C^{K-\ell} W
 \eea
This equation determines $p_{K+R}$ (modulo trace as discussed
above). Moreover tracing both sides gives an equation for
$\delta_K$ \be K\delta_{K} =  - \tr\le( \sum_{j=-K}^{R-1}\mathcal
C^{-K} B_j \mathcal C^{K+j}p_{K+j} \ri) \ , \ K\geq 1\ . \ee This
determines the series $Z$ uniquely. Q.E.D.\par \vskip 5pt
 We can now conclude the proof of Prop. (\ref{lemma2}): to this end we
 gauge transform once more the connection $\pa_q - n B(q)$ by means of
 the formal gauge $ Y:= Z(q) W^{-1} q^G$, namely
 \bea
 \pa_q-nB(q)\mapsto \pa_q - n \tilde D(q)\\
\tilde D(q):= q^{-G} W\overbrace{\le( Z^{-1}B Z  - \frac 1 n
  Z^{-1}Z'\ri)}^{=T'(q)} W^{-1} q^G  -\frac G {q} \ .
 \eea
 The formal series $\tilde D(q)$ enjoys the periodicity
 \be
 \tilde D(\omega q) = \frac  1 \omega \tilde D(q)
 \ee
 Restoring the independent variable to $x$ we have the connection $\pa_x -D(x)$ where
 \be
 D(x) = \frac 1 {x} q\tilde D(q)  = \frac 1 {nx} \sum_{j=0}^{R} t_j \le[
\begin{array}{ccccc}0 & 0 & 0 & 0 & x \\1 & 0 & 0 & 0 & 0 \\0 & 1 & 0
 & 0 & 0 \\0 & 0 & \ddots & 0 & 0 \\0 & 0 & 0 & 1 & 0\end{array}
 \ri]^{j} - \frac G{nx} = \frac 1 {nx} \sum_{j=0}^{R} t_j \mathcal H(x)^j - \frac G{nx}
  \ee
 At this point we can summarize the chain of gauge transformations as
 \bea
 Y^{-1}A\,Y &-& Y^{-1} Y' = D(x)\\
 Y(x) &=& q^{-G} Z(q) W^{-1} q^{G} =  \sum_{j=0}^{\infty} q^{-j} q^{-G}\mathcal C^j z_j q^G=\\
 &=&  \sum_{j=0}^{\infty} \mathcal H^{-j}(x) z_j\ .
 \eea
 Since $z_0=\1$ and $z_j$ are diagonal matrices, a direct inspection shows that the leading coefficient of $Y(x)$ is
 \be
 Y(x) =  \le(\1 +\sum_{j=1}^{n-1} \mathcal N^j z_j + \mathcal O(x^{-1})\ri)\ .
 \ee
 Therefore the formal gauge $Y(x)$ is nonsingular (note also that
  it contains only integer powers of $x$, in spite of the intermediate
  steps). Finally we claim that the constant term in $Y(x)$ is in the
  centralizer of $\mathcal N$\footnote{This does not follow from the
 above expression directly because the diagonal matrices $z_j$ are not
 multiples of the identity.}: indeed by inspecting the leading term
  of $D(x)$ we see that it is the same as the leading term of $A(x)$
  (in the ``canonical'' form \ref{cano}), and hence the constant term
  of $Y(x)$ must commute with it and so commutes also with its
  canonical form, which is $\mathcal N$. We can incorporate $\1 + \sum \mathcal N^j z_j$
  it in the first constant gauge, thus completing the proof.
 Q. E. D. Prop. \ref{lemma2}.\par \vskip 5pt
\br
\label{Casimirs}
For later reference we point out that the upper triangular matrix $U:=
\sum_{j=1}^{n-1}\mathcal N^j z_j$ is uniquely fixed once the connection
is in the canonical form (\ref{cano}). Moreover, from the recursion
relations defining $y_j$'s it is obvious that the entries of $U$ are
rational in $\lambda_1 = t_{rn-1}$ and polynomials in the other
entries of the coefficients of $A(x)$.
\er
 Before proceeding with the analysis of isomonodromic deformations we point out the almost obvious
 \bp[Bare Isomonodromic Deformation]\label{lemmabare}
 Introducing the matrices
 \bea
&& D^{\infty}(x) =  \frac 1 {nx} \sum_{j=0}^{rn} t_j \mathcal H(x)^j - \frac G{nx}\ ,\qquad G:={\rm diag}(0,1,\dots,n-1)\\
 && \mathcal T^{\infty,bare}_J(x) = \frac 1 J \mathcal H(x)^J\ ,\ J=1,\dots, rn\ .
 \eea
(where $\mathcal H(x)$ is the matrix appearing in eq. \ref{shear}) we have the zero-curvature equations
 \bea
 &&\bigg[\frac \pa{\pa t_J} - \mathcal T^{\infty,bare}_J, \frac \pa{\pa t_K} -\mathcal T_K^{\infty,bare}\bigg] =0\label{baredef}\\
 &&\bigg[\frac \pa{\pa t_J} - \mathcal T^{\infty, bare}_J,\pa_x - D^{\infty}(x)\bigg]=0\label{barediff}\ .
 \eea
The kernel of all these connection is spanned by
\be
\Psi^{bare} =   \exp\le(\sum_{J=1}^{rn} \frac {t_J}J\mathcal
  H^J\ri)x^{\frac {t_0-G}n}\ .
\ee
 \ep
{\bf Proof}.
The above connection is nothing but the connection
\be
\pa_q -\sum_{J=0}^{rn}t_J q^{J-1}\Omega^J\ ;\qquad \pa_{t_J} - \frac 1
  J q^J \Omega^J\ ,\ J=1,\dots rn\ ,\ \ q=x^{\frac 1 n}
\ee
--the compatibility of which is absolutely trivial--
written in a different gauge (and variable), namely by gauging with $q^{-G} W$.
The kernel of these connections is trivially
\be
\Phi(q) = {\rm exp} \le(\sum_{J=1}^{rn} \frac {t_J}j q^J\Omega^J +t_0\ln q\ri)
\ee
Therefore we can set (multiplying on the right by the constant
  invertible matrix $W^{-1}$)
\be
\Psi^{bare} = q^{-G} W \,\Phi\,W^{-1} =  q^{-G} W \,\Phi\,W^{-1} q^G
  q^{-G}\ .
 \ee
Since $\mathcal H(x) = q^{-G}W q\Omega W^{-1}q^G$, we find
\be
\Psi^{bare}(x) = \exp\le(\sum_{J=1}^{rn} \frac {t_J}J\mathcal
  H^J\ri)x^{\frac {t_0-G}n}. \ \ \hbox{Q.E.D.}
\ee

Note that in  Prop. (\ref{lemmabare}) the sum extends to $rn$;
  namely $t_{rn}$ is $n\lambda$, where $\lambda$ is the unique
  eigenvalue of the leading coefficient matrix. Among the times, each
  $t_j$, $j\equiv 0 {\rm mod} (n)$ is "trivial" in the sense that it
  could be gauged away by a scalar gauge transformation. However we
  prefer to keep them in view of the general case, where they will not
  be "trivial" anymore. \par
In these considerations we have always placed the pole at $x=\infty$:
  if the pole were at a finite point $x=c$ we should modify some of
  the formulas above in a trivial manner which is left to the
  reader to verify. The analog of Prop. $\ref{lemma2}$ localized at a finite
  point is (we are not considering Fuchsian singularities in this section)

\bp[Prop. \ref{lemma2} for finite poles]\label{lemma2finite}
Let $A(x)$ be a $n\times n$  matrix with a pole at $x=\gamma$ of degree
$r+1\geq 2$  and formal Taylor
series, such that the leading coefficient is $\lambda \1$ plus a nilpotent matrix of rank
$n-1$. Suppose also that the Lidskii coefficient of the subleading
term (in the constant gauge in which $A_r$ is in Jordan canonical
form) does not vanish.
 \be
A(x) = \frac {A_r}{(x-\gamma)^{r+1}} +\frac  {A_{r-1}}{(x-\gamma)^{r}} + \dots
\ee

Then there exists a nonsingular  formal gauge $Y(x)$  of the form
\bea
Y(x) =  C\le(\1 + \mathcal O((x-\gamma))\ri)\ ,
\eea
such that the connection $\pa_x - A(x)$ is formally gauged to the connection $\pa_x - D(x)$ of the form
\bea
D^{\gamma}(x) = \sum_{J=0} ^{rn} \frac {t_J}{n(x-\gamma)} \mathcal
H^J((x-\gamma)^{-1}) - \frac G{n(x-\gamma)}
\\
t_{rn}:= n\lambda\ , \
G:={\rm diag} (0,1,2,\dots,n-1)\\
\mathcal H ((x-\gamma)^{-1})= \le[
\begin{array}{ccccc}0 & 0 & 0 & 0 & (x-\gamma)^{-1}  \\1 & 0 & 0 & 0 & 0 \\0 & 1 & 0
  & 0 & 0 \\0 & 0 & \ddots & 0 & 0 \\0 & 0 & 0 & 1 & 0\end{array}
 \ri]
\eea
\ep
\bp[Bare I-Defs for finite poles]
\label{lemmabarefinite}
 Introducing the matrices
 \bea
D^{\gamma}(x) = \sum_{J=0} ^{rn} \frac {t_J}{n(x-\gamma)} \mathcal H^J((x-\gamma)^{-1}) - \frac G{n(x-\gamma)} \\
 \mathcal T^{\gamma,bare}_{J}(x) := -\frac 1 J \mathcal H((x-\gamma)^{-1})^J\
 ,\ j=1,\dots, rn\\
\mathcal C^{\gamma,bare}(x):= -D^{\gamma}(x)\ ,
 \eea
 we have the zero-curvature equations
 \bea
 &&\bigg[\frac \pa{\pa t_J} - \mathcal T^{\gamma, bare}_J, \frac \pa{\pa t_K} -\mathcal T_K^{\gamma,bare}\bigg] =0\label{baredeffinite}\\
 &&\bigg[\frac \pa{\pa t_J} - \mathcal T^{\gamma,bare}_J,\pa_x - D^{\gamma}(x)\bigg]=0\label{baredifffinite}\\
 &&\bigg[\frac \pa{\pa \gamma} - \mathcal C^{\gamma,bare},\pa_x - D^{\gamma}(x)\bigg]=0\\
 &&\bigg[\frac \pa{\pa \gamma} - \mathcal C^{\gamma,bare}, \frac \pa{\pa t_K} -\mathcal T_K^{\gamma,bare}\bigg] =0
 \eea
The kernel of all these connection is spanned by
\be
\Psi^{bare} =   \exp\le(\sum_{J=1}^{rn} \frac {t_J}J\mathcal
  H^J((x-\gamma)^{-1}) \ri)(x-\gamma)^{\frac {t_0+G}n}\ .
\ee
\ep
We conclude with the remark that the number of isomonodromic times in this situation
 is exactly the same as in the nonresonant case, and in fact this
 still holds for more general Jordan block decompositions: in a sense
 we may consider the cases under analysis ``minimally'' resonant.
\subsection{Isomonodromic Deformations}
\label{IsoDef}
The general situation in which we would like to define a  set of
isomonodromic deformations is that of an arbitrary rational connection
$A(x)$ with leading coefficients at each singularity consisting of (a
conjugacy class of) an arbitrary Jordan form, under the suitable
genericity assumptions on the Lidskii matrix for the subleading
coefficient.

At this point of our discussion the main difficulty is rather
notational than conceptual: in view of Prop. \ref{pro:decomp1} and
Prop. \ref{lemma2} it should be clear that the approach can be
"modular" by reasoning on each block in the local formal gauge at each
of the  singularities of the connection.

It is probably an instructive warm-up exercise to consider the
following simplified but nontrivial situation: let $\pa_x-A(x)$ be a
{\bf polynomial} connection of degree $r-1$ with leading coefficient
consisting of a single Jordan block of size $n$. There is no loss of
generality in assuming that the connection is in the canonical form
(\ref{cano}): this is tantamount requiring that the leading term is
diagonal in the non-resonant case and fixes conveniently the constant
gauge arbitrariness.
More specifically we know from the proof of Prop. \ref{lemma2} that
the coefficients appearing in the canonical form (\ref{cano}) are in
fact the first $n$ times, namely we will have
\be
A(x) =\frac 1 n x^{r-1} \le[
\begin{array}{ccccc}
t_{rn} & t_{rn-1} & \dots&  &t_{rn-n+1}\\
  & t_{rn}&\ddots & & t_{rn-n+2}\\
 & &\ddots &\ddots &\\
&&&t_{rn} & t_{rn-1}\\
&&&&\!\! t_{rn}
\end{array}
\ri] + x^{r-1} \le[\begin{array}{cc}
 & \begin{array}{ccc}
 \phantom{-} & \phantom {-} &\phantom {-}  \\ & { \star} &
 \\ & &
\end{array}
\\
t_{rn}/n &
\end{array}
\ri] + \dots
\label{gaugefixing}
 \ee
where the parameters $t_{rn-j}$, $j=0,\dots, n-1$ are the first $n$
times in the bare form (Prop. \ref{lemmabare}) of the connection.
This can always be achieved -if necessary- by first putting the
leading term in Jordan canonical form, then determining the parameters
$t_{rn-j}$ and then applying a second appropriate constant conjugation.
\bt\label{cool}
Let $A(x)$ be a polynomial of degree $r-1$ with leading
coefficient in the gauge-fixed  form (\ref{gaugefixing}).
There exists a unique formal solution $\Psi' = A\Psi$ of the form
\be
\Psi = \overbrace{\le(\1 + U + \mathcal O(x^{-1})  \ri)}^{=Y(x)}{\exp}
\le[ \sum_{J=1}^{rn} \frac {t_J}J \mathcal H^J(x)\ri]x^{t_0-\frac Gn}
\ ,\ \label{formasympt}
\ee
where $U$ denotes a strictly upper triangular matrix uniquely
determined, rational in $t_{rn-1}$  and polynomial in the remaining coefficients of $A_j$'s, independent of
$x$ and in the centralizer of $\mathcal N$.

We then  define
\be
\mathcal T_J(x):= \frac 1 J \le(Y \mathcal
H^J(x) Y^{-1} \ri)_{+} \ ,
\ee
where $()_{+}$ means the projection
onto nonnegative powers of $x$. With these definitions, the
operators
\be
\pa_x -A(x),\ \pa_j - \mathcal T_J(x)\ ,\ J=1,\dots
rn
\ee
satisfy pairwise zero-curvature conditions, which ensure the
Fr\"obenius integrability of the Pfaffian system
\bea\label{eq:zerocur}
\frac {{\rm d}}{{\rm d}x} \Psi &=&A\Psi\\
\pa_{t_J} \Psi &=& \mathcal T_J \Psi\ .
\eea
The matrix $U$ appearing in the formal gauge is a constant of all the
motions and hence it can be disposed of by an appropriate conjugation
by $\1+U$.
\et
Note that although $U$ is constant and in the centralizer of the
leading coefficient it is not true that it can be set to zero a priori
because setting it to zero is a (although conserved) {\em algebraic
  constraint} on the coefficient of the connection. We should rather
think of $U$ as ``Casimirs'' of the flows.\\
{\bf Proof.}
The property of $U$ being uniquely determined follows from Remark
(\ref{Casimirs}).
 We rewrite the Pfaffian system for $Y$ as follows
\bea
&& \pa_x Y = A\,Y - Y\, D\\
&& \pa_{t_J}Y = \mathcal T_J \,Y - Y\,\mathcal T_J^{bare}\ , \eea
where $D(x)$ and $\mathcal T_J^{bare}$ are as in Prop.
\ref{lemmabare} for \be \Psi^{bare}:=  \exp\le(\sum_{J=1}^{rn}
\frac {t_J}J\mathcal
  H^J\ri)x^{\frac {t_0-G}n}
\ee
is the solution of the Pfaffian system of Prop. \ref{lemmabare}.
The definition of $\mathcal T_j$ implies that
\be
\pa_{t_J}Y Y^{-1} = \mathcal O(x^{-1})\ ,
\ee
namely the upper-triangular matrix $U$ in the constant term is actually
a constant of all the flows;
note that the independent entries of $U$ are $n-1$ since the matrix
commutes with $\mathcal N$ and  depend polynomially on the
(coefficients in $x$ of the) entries of $A$, rationally on $t_{rn-1}$.\\
We now want to verify the zero-curvature conditions; to this end
one computes
\bea
\label{eq:+-}
[\pa_{t_J},\pa_{t_K}]Y\cdot Y^{-1}
&=& \pa_{J}\mathcal T_K-\pa_{K}\mathcal T_J + [\mathcal
T_K,\mathcal T_J] -\overbrace{ \le(  \pa_{J}\mathcal
T_K^{bare}-\pa_{K}\mathcal T_J^{bare} + [\mathcal
T_K^{bare},\mathcal T_J^{bare}]
\ri)}^{\equiv 0} \\[1pt]
[\pa_{t_J},\pa_{x}]Y\cdot Y^{-1} &=& \pa_{J} A -\pa_{x}\mathcal
T_J + [A,\mathcal T_J] -\underbrace{ \le( \pa_{J} D
-\pa_{x}\mathcal T_J^{bare} + [D,\mathcal T_J^{bare}]
\ri)}_{\equiv 0} \ .
\eea

In the above equations the LHS is $\mathcal O(x^{-1})$ since $U$
is a constant of the motions, while the RHS is polynomial.
Therefore both sides vanish identically. Q.E.D.\par\vskip 5pt

Note that
 we could conjugate the
connection $A(x)$ by $\1 + U$ since {\em a posteriori} this is a
 constant gauge of all the flows. However we could not gauge it away
 to begin with, since we did not know {\em a priori} whether it was
 a first integral of the isomonodromic deformations.
\paragraph{Remark.}\label{re:gauge}
We would like to comment about $U$ being a constant in the proof
of theorem \ref{cool}.

The situation is here very similar to the nonresonant case \cite{JMU}:
indeed the deformation equations that we have proposed here are not
the most general that would preserve the leading coefficient at
$\infty$ of the connection $A(x)$. The residual freedom in this case
as well as in the nonresonant case is by gauge action of constant (in
$x$) transformations in the centralizer of $A_r$.

Suppose indeed we have
a general solution $\{A,\mathcal T\}$ to the zero curvature
condition (\ref{eq:zerocur}) such that $\mathcal T$ is a
polynomial. Let their formal solution be given by
\be
 \tilde{\Psi}
= \le(\1 + \tilde{U} + \mathcal O(x^{-1}) \ri){\exp} \le[
\sum_{J=1}^{rn} \frac {t_J} J \mathcal H^J(x)\ri]x^{t_0-\frac Gn} \
\ \ \  .
\ee
Then $\tilde U$ need only be in the centralizer of $A_r$ (i.e. of
$\mathcal N$) but it could otherwise depend on the ``times'' in an
arbitrary analytic way.
 Any such $\tilde{\Psi}$ can be
obtained from multiplying a formal solution (\ref{formasympt}) by
$C=(I+\tilde{U})(I+U)^{-1}$. This has the effect of gauge
transforming the solution $\{A^0,\mathcal T^0\}$ by the gauge $C$,
where $\{A^0,\mathcal T^0\}$ is the solution in which $U$ is
fixed. Therefore any solution to the zero curvature condition can
be obtained by a gauge transformation and there is no lost of
generality in assuming $U$ is a constant.

\subsubsection{Tau function}
\label{taufunctionsimple}
We now turn our attention to a suitable definition of {\bf
  isomonodromic tau function}. To this end we employ the strategy used
  in \cite{BHHP}, namely of expressing the tau function in terms of
  spectral invariants of the connection $A$. This is motivated by the
  fact proven in loc.cit. that the standard Miwa-Jimbo-Ueno definition
  can in fact be expressed in terms of spectral invariants and this
  formulation is more suitable to be generalized to this case.

First of all we make a few simple observations about the formal
gauge $Y$ used in eq. (\ref{formasympt}). Since we have \bea A(x)
&=& Y' Y^{-1}  + Y\le( \frac 1 {nx} \sum_{J=0}^{rn} t_J \mathcal
H(x)^J - \frac G{nx}  \ri)Y^{-1} =\nonumber
 \\
&=&Y\,x^{-\frac G n} W \underbrace{\le(\frac
1{nx}\sum_{J=0}^{rn}t_J  x^{\frac Jn}\Omega^J  - \frac{x^{\frac
Gn} W^{-1}GW x^{-\frac G n}}{nx}\ri)}_{\ds =: \hat D(x)} W^{-1}
x^{\frac Gn} Y^{-1} + Y' Y^{-1}\ , \eea it follows immediately
that the matrix $P(q) := Y\,x^{-\frac
  Gn}W$ is an eigenvector matrix $mod\ q^{-rn-1+n}$  where $q=x^{\frac
  1 n}$. It also follows that the expansion of the eigenvalues $y_j$ of $A$ coincides with the expansion of the eigenvalues of the bare system $\hat D(x)$.
This leads to
\be
J \,\mathcal T_J =\le(Y \mathcal H^J Y^{-1}\ri)_+ =\le( x^{\frac Jn}Y x^{-\frac Gn}
W\Omega^JW^{-1}x^{\frac Gn} \ri)_+ = \le(q^J P \Omega^J P^{-1}\ri)_+ +
\hbox{constant}\ ,
\ee
where the plus in the subscript always denotes the polynomial part in
$x$. The constant appearing in the RHS is actually present only for
$J\geq rn-n+1$, namely for the deformations along the highest $n$
times; in any case the specific form of this constant (w.r.t. $x$)
matrix is irrelevant for our purposes.\\
Note that  the expression in the RHS is nothing but a combination
of the spectral projectors of the matrix $A(x)$; if we denote by
$E_a$ the diagonal elementary matrix we have \be P\Omega^JP^{-1} =
\sum_{\sigma=1}^n \omega^{J(\sigma-1)} P E_\sigma P^{-1} =
\sum_{\sigma=1}^n \omega^{J(a-1)}\Pi_a\ , \ee where $\Pi_a$ is the
rank-one projector on the eigenspace of the eigenvalue $y_a(q)$
(part of the cyclically permuted $n$-tuplet). By the properties of
the spectrum of $A$ we have \be \Pi_\sigma(\omega q) =
\Pi_{\sigma+1}(q). \ee This ensures that the expressions \be
q^J\sum_{\sigma=1}^n \omega^{J(\sigma-1)} \Pi_\sigma(q)\ \ee
contain in fact only {\bf integer powers of $x=q^n$}.
We can write the spectral projectors by the classical formula
\be
\Pi_\sigma(q) = \frac 1{\tr(\widetilde {A-y_\sigma})} \widetilde {A-y_\sigma}\ ,
\ee
where the tilde denotes the classical adjoint (the matrix of
cofactors).\par
Note that once we have fixed the determination of the root $q=x^{\frac
  1n}$  and of $t_{rn-1}=(A_r)^{1/n}$,  there is a unique eigenvalue which admits the (in fact
convergent) Puiseux series expansion
\be
 y_1(q) = \frac 1 n \sum_{J=1}^{rn} t_J q^{J-n} + \frac {\tilde t_0} {nx} +  \frac 1 n \sum_{K=1}^{\infty} K H_K q^{-n-K}\ ,
\ee
where this formula has defined symbols for the coefficients of the
negative powers of the  expansion.
From an algebro-geometric point of view the cyclicity of the
eigenvalues near $x=\infty$ means that the spectral curve has a
branch-point of order $n$ in $\mathbb P^1\times \mathbb P^1$ at $y=\infty$ even
after desingularization; put otherwise the local parameter at near
$x=\infty$ is $q=x^{\frac 1n}$.
This means that we have
\bea
t_J &=& -\res{x=\infty}\le(\sum_{a=0}^{n-1} x^{-J/n}\omega^{Ja} y_{a+1}(q)\ri) {\rm d}x=
\res{\zeta_\infty}x^{-\frac Jn} y{\rm d}x
\ ,\qquad  1\leq J
\leq rn\\
\tilde t_0 &=& -\res{x=\infty}\le(\sum_{a=1}^n y_{a}(q) \ri){\rm d}x   = -\res{x=\infty}\tr(A){\rm d}x= t_0 -\frac {\tr(G)}{n} = t_0-\frac {n-1}2\label{strange}\\
H_J &=& -\frac 1 J \res{x=\infty}\le(\sum_{a=0}^{n-1}x^{J/n}\omega^{Ja} y_a(q)\ri){\rm d}x=
\frac 1 J\res{\zeta_\infty} x^{\frac Jn} y{\rm d}x\ .
\eea
We remark that the first residue-formulas for $t_J$ and $H_J$ are
taken on the $x$-plane (the ``base-curve'') and they make sense
because the expressions in the brackets contain only integer powers of
$x$ due to the cyclicity of the eigenvalues: the second
residue-formulas are taken on the spectral curve around the point
$\zeta_\infty$ (with local parameter $x^{-1/n}$) which projects down to $x=\infty$. Since the positive
orientation of a small circle around $\zeta_\infty$ is the opposite
than the positive (counterclockwise) orientation in the $x$-plane,
this explains the difference in sign.
Note also that the coefficient $t_0$ of the bare system is not the
residue of  the eigenvalue but it is shifted due to the fact that the
term $G/x$ contributes to the residue (\ref{strange}).
We are now ready to define the tau function for this guide-example
\bp
\label{pro:tausimple}
The following differential is closed
\be
{\rm d}\ln \tau = \sum_{K=1}^{rn} {H_K}{\rm d}t_K\ .
\ee
\ep
Before entering in the (easy) proof let us comment on the motivation and shape of the
formula.
The reason for the name of this differential as ``isomonodromic tau
function'' resides in the fact that for the nonresonant case
\cite{BHHP} a similar formula involving residues of the spectral
differential $y{\rm d}x$ on the spectral curve of the connection is
shown to coincide with the definition in \cite{JMU}, which is of  different nature and --on the face of it-- not a spectral invariant of the connection.
Secondly, this sort of expressions appear throughout the literature of
dispersionless integrable hierarchies, Seiberg-Witten models etc.,
we cannot find in the literature any statement or implication that the
Miwa-Jimbo-Ueno tau function is defined  by or equivalent to  this
ubiquitous formula.

On a more technical point we also remark that the
most similar setting is that of the so-called universal Whitham
hierarchy \cite{krikko}, where the data are an algebraic curve and a
meromorphic differential plus some decorative data of local parameters
near punctures.

While the main characters appear similar (we have a spectral curve, we
have punctures and local parameters and a meromorphic differential
$y{\rm d}x$) there are some significant distinctions regarding the
coordinates on the phase space. In fact in the Whitham setting the
periods of the differential $y{\rm d}x$ would be treated as
coordinates on the phase space independent of the other ``times''. In
this case the periods of the differential $y{\rm d}x$ are {\bf not
  independent} and in fact mainly uncontrollable. What is fixed
instead are the parameters of formal monodromy and the Stokes'
parameters, which cannot be recovered by inspection of the spectral
curve alone.\\
{\bf Proof of Prop. \ref{pro:tausimple}.}\\
We compute the closure of the differential
\be
- \pa_{t_J}H_K = \frac 1 K\res{\zeta_\infty} x^{K/n} \pa_{t_J}
y{\rm d}x =  \frac 1 {J} \le(\sum_{a=1}^n \res{q=\infty}
x^{\frac Kn}\omega^{K(a-1)}\pa_{t_J} y_a(x)\ri){\rm d}x\ .
\ee
In order to compute the variation of the eigenvalue we recall the
classical formula describing the variation of a simple eigenvalue $y$
of a matrix $A$
under an infinitesimal deformation $A\mapsto A+\delta A$
\be
\delta y = \frac 1{\tr(\widetilde{A-y})} \tr\le(\widetilde{A-y} \,\,\delta
A\ri) \ .
\ee

In our case we should use the formula with $\delta A = \mathcal T_J' +
[\mathcal T_J,A]$ because on the path of integration the eigenvalue
is ``uniformly'' simple.
 We thus obtain
\bea
-\pa_{t_J}H_K = \frac 1 {K} \res{x=\infty}\le(\sum_{b=1}^n
\omega^{K(b-1)} x^{\frac K n } \tr\le(
\frac{(\widetilde{A-y_b})(x)} {\tr (\widetilde{A-y_b)}(x) } \le(\pa_x
\mathcal T_J + [\mathcal T_J,A]\ri)\ri)\ri){\rm d}x  = \\
= \res{x=\infty}\sum_{b=1}^n \omega^{K(b-1)}  \frac 1{JK} x^{\frac K n} \tr\le(
\Pi_b(x) \frac {{\rm
    d}}{{\rm d}x}\le(\sum_{a=1}^n x^{\frac Jn} \omega^{J(a-1)} \Pi_a(x)\ri)_+ \ri) {\rm d}x = \\
    = \frac 1{JK}  \res{x=\infty} \res{z=\infty} \le(\sum_{b,a=1}^n \omega^{K(b-1) + J(a-1)}
  \frac{ x^{\frac K n}z^{\frac J n}}{(x-z)^2}
    \tr\le(\Pi_b(q)  \Pi_a(p)\ri)\ri)  {\rm d}z\,{\rm d}x \ .
 \eea
 At this point the formula is almost obviously symmetric: attention  is to be paid as to whether the order of residues can be
 interchanged. To see this, consider the kernel
 \be
\Omega_{ba}(x,z) := \frac 1{(x-z)^2} \tr\le(\Pi_b(x)\Pi_a(z) \ri){\rm d}x {\rm d}z\ .
 \ee
 To conclude that the order of residues can be interchanged it is
 sufficient to show that there is no residue  on the diagonal
 $x=z$. To this end we note that
 \be
 \tr(\Pi_a(x)\Pi_b(z) )= \delta_{ab} + \mathcal O((x-z)^2).
 \ee
 Indeed the fact that $\tr(\Pi_a(z)\Pi_b(x)) = \delta_{ab}$ is nothing
 but a statement of simplicity of eigenvalues ($x\neq \infty$): by
 continuity it holds also at $x=\infty$. Next, we note that if $C(x)$
 is a locally differentiable matrix of eigenvectors, we have
 \bea
 \frac {\tr(\Pi_a(x)\Pi_b(z))}{x-z} &=& \tr (\Pi_a(x)\Pi_b'(x)) + \mathcal O(x-z) = \nonumber\\
 &=&
 \tr\le(\Pi_a[C'C^{-1}, \Pi_b]\ri) + \mathcal O(x-z)= \nonumber\\
 &=& \tr(C'C^{-1} [\Pi_a,\Pi_b]) +\mathcal O(x-z)= \mathcal O(x-z) \ ,
 \eea
 where we have used the cyclicity of trace. This proves that the kernel
 $\Omega_{ab}(x,z)$ has only a double pole without residue on the
 diagonal (and only for $a=b$) and this is sufficient to have independence of the order of
 the residues. Q.E.D.\par\vskip 5pt
Note that in fact the kernel introduced in the proof has some interesting properties because it is
analytic on the whole spectral curve except on the diagonal where it
has a residueless normalized double pole. This is almost the same as
the fundamental bidifferential of the spectral curve, the only
shortcoming being the absence of a definite normalization around
an isotropic basis in the homology of the curve.\par\vskip 5pt
Before taking on the case of a general rational connection, we
highlight the overall logic used to construct the Pfaffian system,
prove Fr\"obenius integrability and construct the tau function:
\begin{enumerate}
\item  we start with the {\bf bare} zero-curvature equations in Prop. \ref{lemmabare}
\item we have {\bf dressed} by the formal {\bf nonsingular} gauge $Y$
\item we have defined the tau function in terms of residues of a canonical differential on the spectral curve.
 \end{enumerate}
The key of the proof of zero-curvature is the usual argument that the
curvature of the Pfaffian system is on one hand {\em a priori} a
polynomial expression in $x$ whereas on the other side it is also a
Laurent series, hence concluding that it must identically vanish: this
argument -of course- hinges on the fact that the formal gauge $Y$ is
nonsingular.
The deviation from the usual situation is not in the way we prove the
compatibility, but rather in the precise form of the bare system of
equations which was not known beforehand and  which in the nonresonant
case is in diagonal form.

\section{Fuchsian resonant case}\label{se:fuch}
Isomonodromic deformations of resonant Fuchsian system in full
generality have been addressed in \cite{bolibruch} so that here we
mainly collect some known facts for the reader's convenience.
Consider a formal Fuchsian singularity (at $z=0$)
\bea
F(z) = \frac {\Lambda}z + \sum_{j\geq 0}{F_j}z^j\ .
\eea
We assume without loss of generality that $\Lambda$ is in Jordan
canonical form. Since the notion of resonance for (formal) Fuchsian
singularities is that two eigenvalues differ by a {\bf nonzero
  integer}, we split $\Lambda$ into blocks of (Jordan blocks of)
eigenvalues differing only by integers (a {\bf bouquet}). It is not difficult to show
that there is a formally analytic gauge in which the connection is
completely split into block diagonal form according to the
decomposition into bouquets. Hence we can assume that $\Lambda$ has
only eigenvalues differing by integers (i.e. consists of only one
bouquet of eigenvalues). By a scalar gauge transformation we can
actually shift the eigenvalue to zero.\par
At the end of these simplifications we have a connection in which
$\Lambda$ is in Jordan canonical form with only integer eigenvalues
which we assume in decreasing order
\be
n_0 > n_1>n_2 >\dots >n_K\ .
\ee
Each eigenvalue has a certain algebraic multiplicity and a certain
geometrical multiplicity (i.e. the rank of the eigenspace), which is of no particular interest to us.

We now consider a formal analytic gauge $Y(z) = \sum_{j\geq 0} Y_j
z^j$ and seek the most ``canonical'' form under formal gauge
equivalence
\be
Y' = F Y - Y B\ ,\qquad B  = \frac \Lambda z  + \sum_{j \geq 0} B_j z^j \ ,   \ \ \ Y=\1 + \mathcal O(z)
\ee
Our goal is to have $B$ as simple as possible. Writing out the coefficients of the power $z^{k-1}$ we have
\be
k Y_k = [\Lambda,Y_k] + F_k- B_k + \sum_{j=1}^{k-1} \le(F_jY_{k-j} - Y_{k-j} B_j\ri)
\ee
We rewrite this as
\be
(k\, {\rm Id} - ad_\Lambda)Y_k  = F_k - B_k + \sum_{j=1}^{k-1} \le(F_jY_{k-j} - Y_{k-j} B_j\ri)\ .
\label{recfuch}
\ee
The linear operator $\mathcal L_k:= (k\, {\rm Id} - ad_\Lambda)$ on
the space of matrices $n\times n$ is invertible provided that $k$ is
not in the spectrum of  $ad_{\Lambda}$, i.e. provided that no pair of
eigenvalues of $\Lambda$ differ by $k$. If $\mathcal L_k$ is
invertible we can impose $B_k=0$ since the solvability of the
recurrence relation in terms of $Y_k$ is guaranteed. If $\mathcal L_k$
is not invertible  then $B_k$ must be chosen so that the RHS of
(\ref{recfuch}) is in the image of $\mathcal L_k$. It is not difficult
to see that the image of $\mathcal L_k $ consists of arbitrary
matrices with a zero block in the $(j,\ell)$ block such that
$n_j-n_\ell = k$. Therefore $B_k$ can be chosen to be zero in the
complement of that block and it is then uniquely determined by
(\ref{recfuch}) itself. By finiteness of the number of eigenvalues of
$\Lambda$ we can assure that $\mathcal L_k $ is invertible for $k$
large enough, namely that only a finite number of $B_k$ may need to be
chosen nonzero. At the end of this procedure we always obtain a
connection in the form
\bea
B = \frac 1 z \le[
\begin{array}{c|c|c|c|c|c}
n_0 + \triangledown  & \star z^{n_0-n_1} & \star z^{n_0-n_2} & \star z^{n_0-n_3} & \dots
 &\star z^{n0-n_K}\\
\hline
0 & n_1\1  + \triangledown & \star z^{n_1-n_2} & \star z^{n_1-n_3} &
 \dots & \star z^{n_1-n_K}\\
\hline
0 & 0& n_2\1 + \triangledown & \star z^{n_2-n_3} & \star z^{n_2-n_4}
 &\dots\\
\hline
& & & \ddots & &\\
\hline
&&&&n_{K-1}\1 + \triangledown& \star z^{n_{K-1}-n_K}\\
\hline
& & & & &n_K\1 + \triangledown
\end{array}
\ri] \label{canon}
\eea
 where $\triangledown$ denotes a nilpotent matrix in
Jordan canonical form and $\star$ denote the only possibly nonzero
coefficients and {\bf constant} in $z$. Each diagonal block in the
above decomposition has dimension equal to the algebraic
multiplicity of the corresponding integer eigenvalue of $\Lambda$. The block
diagonal shearing
\be
z^G:=(z^{n_0}\1,z^{n_1}\1,z^{n_2}\1,\dots)
\ee
(here each identity is of the appropriate dimension)
recasts the connection to a nonresonant one with the only
eigenvalue zero of the simple form
\be
z^{-G} B z^{G} - \frac G z = \frac {T}z\ ,\ \ T:=  \le[
\begin{array}{c|c|c|c|c|c}
\triangledown  & \star & \star & \star & \dots
 &\star \\
\hline
0 & \triangledown & \star & \star &
 \dots &\star \\
\hline
0 & 0&\triangledown & \star & \star
 &\dots\\
\hline
& & & \ddots & &\\
\hline
&&&&\triangledown&\star\\
\hline
& & & & & \triangledown
\end{array}
\ri] \label{canoconst}
\ee
Note that $T$ is constant.
 The $\star$'s in (\ref{canon}) and (\ref{canoconst}) represent the same coefficients and  are defined up to action of
constant gauge transformations in block diagonal form, each
nonzero block of which consists of the centralizers of the
diagonal blocks of $F$. The orbit under this centralizer group
is what defines  the local  monodromy and hence must be
preserved by the isomonodromic deformation.\\
The solution of last system is
\be
\Phi^{bare} = z^{T}.
\ee
In other words a (formal) solution of the original system is
\be
\Psi = Y(z) z^{G}z^{T} \ ,\label{psiform}
\ee
where $G$ is the diagonal matrix of integers used in the shearing
(inducing the grading) and in general does not commute with $T$.
Since $T$ is upper (semi)triangular it can be put in Jordan canonical form $T_{can}$ by an upper triangular matrix $P$;
\be
\Psi = Y z^G P z^{T_{can}} P^{-1}
\ee
Since $z^{G}P z^{-G}$ is analytically invertible, it can be reabsorbed
in the definition of $Y$, so that without loss of generality we can
always assume the formal solution (\ref{psiform}) to have $T$ in
Jordan canonical form and hence we will denote $T$ by $J$ in the sequel.
At this point the residual arbitrariness of this (formal) solution is
by multiplication on the right by a $z$-independent matrix  $S$ in the
centralizer of $J$ and such that $z^G S z^{-G}$ is analytic; this
implies that $S$ must be at the same time in the centralizer of $J$
and in block upper-triangular form according to the minimal
decomposition of $G$ into blocks which are multiple of the identity
matrix.

To put it in a different way, $S$ must be in the intersection of the
positive root spaces of $Ad_{z^G}$ with the centralizer of $J$.

Summarizing this discussion and restoring the generality of all the steps we have obtained the
\bp[Formal solution for resonant Fuchsian singularities]
\label{pro:fuch}
Let $A = \frac {A_0}z + \mathcal O(1)$ be  the matrix of a Fuchsian
singularity at $z=0$. Then there exists a (formal) solution $\Psi$ of
the form
\be
\Psi = Y(z) z^G z^J
\ee
where
\begin{enumerate}
\item $Y(z)$ is analytically (formally) invertible at $z=0$
\item $J$ is in Jordan canonical form with distinct  eigenvalues with real part in the interval $[0,1)$
\item $G$ is an integer valued diagonal matrix such that  --within each
  block corresponding to the same eigenvalue of $T$-- the integers form
  a weakly decreasing sequence which distinguishes the eigenvalues of
  the same bouquet.
\item The spectrum of $A_0$ coincides with the spectrum of $G+J$ and moreover  $A_0 = Y_0 (G+J){Y_0}^{-1}$
\end{enumerate}
Such solution is unique up to ordering of the eigenvalues of $J$ and
by multiplication on the right by a $z$-independent matrix lying in
the intersection of the centralizer of $J$ and the nonnegative root
subspace of $Ad_{z^G}$\footnote{The operator $Ad_{z^G}$ splits
  $GL(n,\C)$ into eigenspaces with eigenvalues $z^{G_{ii}-G_{jj}}$;
  the direct sum of the subspaces with nonnegative exponent for this
  eigenvalue is our definition of  nonnegative root subspace. For
  example if the entries of $G$ are strictly decreasing then the
  nonnegative root subspaces are the upper semitriangular matrices.}.
  Moreover the gauge $Y(z)$ is actually a convergent series if $A(z)$ is convergent in a punctured disk around the regular singularity \cite{wasow}.
\ep

\subsection{Isomonodromic deformation of resonant Fuchsian singularities}
Here we are rephrasing part of the content of \cite{bolibruch}.
Suppose we have an isomonodromic family of resonant Fuchsian
connections with poles at points $\gamma_j$, $j=1,\dots $  and let
$\Phi(z;{\boldsymbol \gamma})$ be the fundamental solution of the family. At each
pole $\gamma_j$  and by virtue of the previous discussion culminated
in Prop. \ref{pro:fuch} there is a $z$-independent  nonsingular matrix
$C_j$ for which
\be
\Phi =  Y_j (z) (z-\gamma_j)^{G_j} (z-\gamma_j)^{J_j}C_j\ .
\ee
The matrix $C_j$ is defined modulo the group described in Prop. \ref{pro:fuch}.
The monodromy is then
\be
M_j = C_j^{-1} {\rm e}^{2i\pi J_j} C_j\ .
\ee
We see that under a monodromy preserving deformation the matrices
$C_j$ can vary arbitrarily by left multiplication of a matrix in the centralizer of $J_j$ (which is the
same as  the  centralizer of ${\rm e}^{2i\pi J_j}$ since the
eigenvalues of $J_j$ by construction do not differ by integers).\par

\begin{figure}
\begin{center}
\setlength{\unitlength}{3000sp}%
\begingroup\makeatletter\ifx\SetFigFont\undefined%
\gdef\SetFigFont#1#2#3#4#5{%
  \reset@font\fontsize{#1}{#2pt}%
  \fontfamily{#3}\fontseries{#4}\fontshape{#5}%
  \selectfont}%
\fi\endgroup%
\begin{picture}(2124,2124)(1189,-2473)
\thinlines
{\color[rgb]{0,0,0}\put(1201,-828){\line( 1, 0){2100}}
}%
{\color[rgb]{0,0,0}\put(1668,-361){\line( 0,-1){2100}}
}%
{\color[rgb]{0,0,0}\put(1201,-1528){\line( 1, 0){2100}}
}%
{\color[rgb]{0,0,0}\put(2368,-361){\line( 0,-1){2100}}
}%
{\color[rgb]{0,0,0}\put(1201,-2461){\framebox(2100,2100){}}
}%
\thicklines
{\color[rgb]{0,0,0}\put(1376,-419){\line( 1,-1){233.500}}
}%
{\color[rgb]{0,0,0}\put(1843,-886){\line( 1,-1){466.500}}
}%
{\color[rgb]{0,0,0}\put(2543,-1586){\line( 1,-1){700}}
}%
\end{picture}%
\hspace{3cm}
\setlength{\unitlength}{3000sp}%
\begingroup\makeatletter\ifx\SetFigFont\undefined%
\gdef\SetFigFont#1#2#3#4#5{%
  \reset@font\fontsize{#1}{#2pt}%
  \fontfamily{#3}\fontseries{#4}\fontshape{#5}%
  \selectfont}%
\fi\endgroup%
\begin{picture}(2124,2124)(1189,-2473)
\thicklines
{\color[rgb]{0,0,0}\put(1376,-419){\line( 1,-1){233.500}}
}%
{\color[rgb]{0,0,0}\multiput(1494,-419)(6.07895,-6.07895){20}{\makebox(6.6667,10.0000)
    {\SetFigFont{7}{8.4}{\rmdefault}{\mddefault}{\updefault}.}}
}%
{\color[rgb]{0,0,0}\put(1608,-419){\makebox(6.6667,10.0000){\SetFigFont{10}{12}{\rmdefault}{\mddefault}{\updefault}.}}
}%
{\color[rgb]{0,0,0}\put(1259,-419){\line( 1,-1){349}}
}%
{\color[rgb]{0,0,0}\put(1843,-886){\line( 1,-1){466.500}}
}%
{\color[rgb]{0,0,0}\put(1959,-886){\line( 1,-1){349.500}}
}%
{\color[rgb]{0,0,0}\put(2076,-886){\line( 1,-1){232.500}}
}%
{\color[rgb]{0,0,0}\multiput(2194,-886)(6.07895,-6.07895){20}{\makebox(6.6667,10.0000){\SetFigFont{7}{8.4}{\rmdefault}
      {\mddefault}{\updefault}.}}
}%
{\color[rgb]{0,0,0}\put(2308,-886){\makebox(6.6667,10.0000){\SetFigFont{10}{12}{\rmdefault}{\mddefault}{\updefault}.}}
}%
{\color[rgb]{0,0,0}\put(1726,-886){\line( 1,-1){582}}
}%
{\color[rgb]{0,0,0}\put(2543,-1586){\line( 1,-1){700}}
}%
{\color[rgb]{0,0,0}\put(3243,-2168){\line(-1, 1){583}}
}%
{\color[rgb]{0,0,0}\put(2776,-1586){\line( 1,-1){467.500}}
}%
{\color[rgb]{0,0,0}\put(2894,-1586){\line( 1,-1){349.500}}
}%
{\color[rgb]{0,0,0}\put(3008,-1586){\line( 1,-1){234}}
}%
{\color[rgb]{0,0,0}\multiput(3126,-1586)(6.15789,-6.15789){20}{\makebox(6.6667,10.0000){\SetFigFont{7}{8.4}{\rmdefault}
      {\mddefault}{\updefault}.}}
}%
{\color[rgb]{0,0,0}\put(3243,-1586){\makebox(6.6667,10.0000){\SetFigFont{10}{12}{\rmdefault}{\mddefault}{\updefault}.}}
}%
{\color[rgb]{0,0,0}\put(2426,-1586){\line( 1,-1){817}}
}%
{\color[rgb]{0,0,0}\put(2776,-886){\line( 1,-1){467.500}}
}%
{\color[rgb]{0,0,0}\put(2894,-886){\line( 1,-1){349.500}}
}%
{\color[rgb]{0,0,0}\put(3008,-886){\line( 1,-1){234}}
}%
{\color[rgb]{0,0,0}\multiput(3126,-886)(6.15789,-6.15789){20}{\makebox(6.6667,10.0000){\SetFigFont{7}{8.4}{\rmdefault}
      {\mddefault}{\updefault}.}}
}%
{\color[rgb]{0,0,0}\put(3243,-886){\makebox(6.6667,10.0000){\SetFigFont{10}{12}{\rmdefault}{\mddefault}{\updefault}.}}
}%
{\color[rgb]{0,0,0}\put(2659,-886){\line( 1,-1){583}}
}%
{\color[rgb]{0,0,0}\put(1843,-1586){\line( 1,-1){466.500}}
}%
{\color[rgb]{0,0,0}\put(1959,-1586){\line( 1,-1){349.500}}
}%
{\color[rgb]{0,0,0}\put(2076,-1586){\line( 1,-1){232.500}}
}%
{\color[rgb]{0,0,0}\multiput(2194,-1586)(6.07895,-6.07895){20}{\makebox(6.6667,10.0000){\SetFigFont{7}{8.4}{\rmdefault}
      {\mddefault}{\updefault}.}}
}%
{\color[rgb]{0,0,0}\put(2308,-1586){\makebox(6.6667,10.0000){\SetFigFont{10}{12}{\rmdefault}{\mddefault}{\updefault}.}}
}%
{\color[rgb]{0,0,0}\put(1726,-1586){\line( 1,-1){582}}
}%
{\color[rgb]{0,0,0}\put(1376,-1586){\line( 1,-1){232.500}}
}%
{\color[rgb]{0,0,0}\multiput(1494,-1586)(6.07895,-6.07895){20}{\makebox(6.6667,10.0000){\SetFigFont{7}{8.4}{\rmdefault}
      {\mddefault}{\updefault}.}}
}%
{\color[rgb]{0,0,0}\put(1608,-1586){\makebox(6.6667,10.0000){\SetFigFont{10}{12}{\rmdefault}{\mddefault}{\updefault}.}}
}%
{\color[rgb]{0,0,0}\put(1259,-1586){\line( 1,-1){349.500}}
}%
{\color[rgb]{0,0,0}\put(2076,-419){\line( 1,-1){233.500}}
}%
{\color[rgb]{0,0,0}\multiput(2194,-419)(6.07895,-6.07895){20}{\makebox(6.6667,10.0000){\SetFigFont{7}{8.4}{\rmdefault}
      {\mddefault}{\updefault}.}}
}%
{\color[rgb]{0,0,0}\put(2308,-419){\makebox(6.6667,10.0000){\SetFigFont{10}{12}{\rmdefault}{\mddefault}{\updefault}.}}
}%
{\color[rgb]{0,0,0}\put(1959,-419){\line( 1,-1){349}}
}%
{\color[rgb]{0,0,0}\put(3008,-419){\line( 1,-1){235}}
}%
{\color[rgb]{0,0,0}\multiput(3126,-419)(6.15789,-6.15789){20}{\makebox(6.6667,10.0000){\SetFigFont{7}{8.4}{\rmdefault}
      {\mddefault}{\updefault}.}}
}%
{\color[rgb]{0,0,0}\put(3243,-419){\makebox(6.6667,10.0000){\SetFigFont{10}{12}{\rmdefault}{\mddefault}{\updefault}.}}
}%
{\color[rgb]{0,0,0}\put(2894,-419){\line( 1,-1){349}}
}%
{\color[rgb]{0,0,0}\put(1376,-886){\line( 1,-1){232.500}}
}%
{\color[rgb]{0,0,0}\multiput(1494,-886)(6.07895,-6.07895){20}{\makebox(6.6667,10.0000){\SetFigFont{7}{8.4}{\rmdefault}
      {\mddefault}{\updefault}.}}
}%
{\color[rgb]{0,0,0}\put(1608,-886){\makebox(6.6667,10.0000){\SetFigFont{10}{12}{\rmdefault}{\mddefault}{\updefault}.}}
}%
{\color[rgb]{0,0,0}\put(1259,-886){\line( 1,-1){349.500}}
}%
\thinlines
{\color[rgb]{0,0,0}\put(1201,-828){\line( 1, 0){2100}}
}%
{\color[rgb]{0,0,0}\put(1668,-361){\line( 0,-1){2100}}
}%
{\color[rgb]{0,0,0}\put(1201,-1528){\line( 1, 0){2100}}
}%
{\color[rgb]{0,0,0}\put(2368,-361){\line( 0,-1){2100}}
}%
{\color[rgb]{0,0,0}\put(1201,-2461){\framebox(2100,2100){}}
}%

\end{picture}%
\caption{\label{centralizer}
A nilpotent matrix in Jordan canonical form and the shape of its centralizer: the oblique segments
  represent entries with the same numerical value \cite{arnold}.}
\end{center}
\end{figure}
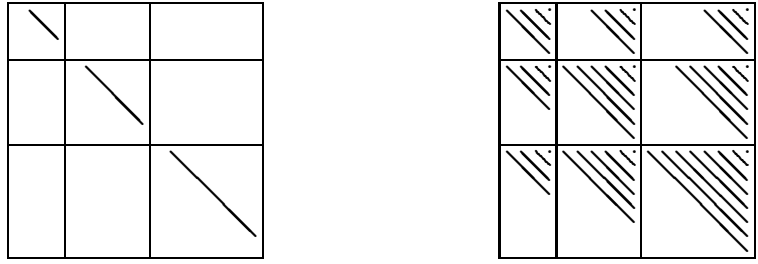

Under a continuous monodromy preserving deformation we have near each pole $\gamma_j$
\be
\dot \Phi \Phi^{-1} = \dot \gamma_j\,Y_j \frac {G_j+J_j}{z-\gamma_j} {Y_j}^{-1}
+ Y_j (z-\gamma_j)^{G_j} \dot C_j {C_j}^{-1} (z-\gamma_j)^{-G_j} {Y_j}^{-1} + \dot Y_{j}Y_{j}^{-1}\ .
\ee
The last term is analytic at $\gamma_{j}$.
The first term has a simple pole and is the standard term in
Schlesinger deformations. The second term may have poles of higher
order: indeed $\dot C_j C_j$ needs only to belong to the centralizer
Lie algebra of $J_j$ but not necessarily to the centralizer of
$G_j$ nor to its nonnegative root subspace. A case in which the second term is certainly absent is
when the centralizer of $J_j$ is the Abelian algebra of diagonal
matrices, which corresponds to the case of nonresonant Fuchsian
singularities; another case is when $J_j$ has only one Jordan block,
for in that case the centralizer is upper triangular and hence
certainly in the nonnegative root space of $Ad_{z^G}$. In case $J_j$ contains more than one irreducible block with the same
eigenvalue then the centralizer is not upper triangular (see  Figure \ref{centralizer}
for representations of such a centralizer) and
hence conjugation by $(z-\gamma_j)^{G_j}$ may have poles  of higher degree
(at most the index of resonance, i.e., the maximum integer difference between two eigenvalues).\par \vskip 6pt
This situation is the most general as explained in
\cite{bolibruch}. We can regard the dependence of $C_j$ on the
deformations as ``pure gauge'' in the sense that it is arbitrarily
defined and can always be
disposed of by a {\bf rational} gauge equivalence which {\em does not
  move the position of the poles}, or -which is the
same- by solving a certain Riemann--Hilbert problem.

Suppose indeed that $\Phi, \tilde \Phi$, are two isomonodromic
families of a resonant Fuchsian connection  with the same monodromy
representation $\{M_j\}$ {\bf and} with same residual spectrum at the Fuchsian singularities
\be
Sp(\res{z=\gamma_j}\Phi'\,\Phi^{-1}) = Sp(\res{z=\gamma_j}\tilde \Phi' \,\tilde \Phi^{-1})\ ,\ \forall j\ .\label{resspec}
\ee
This implies that the matrices $J_j$ (the Jordan form of the monodromies)  and $G_j$ are the same for the two families.
 This also implies that
\be
G (z):= \tilde  \Phi(z) \Phi^{-1}(z)
\ee
 is a matrix function defined on the punctured plane  and single-valued. One sees from the asymptotic
 representation of $\tilde \Phi$ and $\Phi$ that
 \be
 G(z)  =  \tilde Y_j (z-\gamma_j)^{G_j} \tilde C_j C_j^{-1} (z-\gamma_j)^{-G_j} Y_j + \mathcal O(z-\gamma_j)
 \ee
where we have used that $H_j:= \tilde C_j C_j^{-1}$ is in the centralizer
group of $J_j$ because $C_j^{-1} {\rm e}^{2i\pi J_j} C_j = M_j =
\tilde M_j  = \tilde C_j^{-1} {\rm e}^{2i\pi J_j} \tilde C_j$.
This shows that $G(z)$ has at worst poles at the $\gamma_j$'s and hence it is rational.

We can think of $G(z)$ as the solution of the following RH problem
\bea
G_{+}(z) =  G_{-} (z) \, Y_j \,(z-\gamma_j)^{G_j} H_j
(z-\gamma_j)^{-G_j} {Y_j}^{-1} \ , \ \ z\in \{|z-\gamma_j| =
\epsilon \} \label{RHG}
\eea
where $G_-(z)$ is analytically invertible in each disk and equal to $\widetilde Y_j(z){Y_j}^{-1}(z)$ (recall that both $\widetilde Y_j$ and $Y_j$ are actual convergent series around $\gamma_j$ \cite{wasow}). Since the jump matrices are analytic in each punctured small disk, the ``exterior'' part of the solution of this RH problem can be analytically extended to a rational function $G(z)$ in the punctured plane $\C P^1\setminus \cup \{\gamma_j\}$.

Vice-versa suppose that given an isomonodromic family $\Phi$ we want to
pass to another isomonodromic family $\tilde \Phi$ with the same
monodromy matrices $M_j$ and same residual spectrum (\ref{resspec}).
To this end we should find a solution to the  Riemann--Hilbert problem (\ref{RHG}) with preassigned arbitrary
matrices $H_j$ depending on the  deformation
parameters (but independent of $z$) and in the  centralizer
group of $J_j$.
The ``exterior'' part $G_+$ of the  solution of the RH problem defines by analytical continuation a rational function $G=G_+$ which transforms the family $\Phi$ in a family $\tilde \Phi$ with $\tilde C_j= H_j C_j$. The solvability of this RH problem can be assured by the argument which will be used later in the more general setting of Thm. \ref{thm:isogen} (see Remark \ref{RHrem}).

This discussion means that for resonant Schlesinger systems the
Schlesinger equations are still consistent and any other isomonodromic
family is obtained from a solution of Schlesinger equations by a
rational gauge equivalence constructed from a solution of
(\ref{RHG}).

\section{General Case}
We now address the full-fledged general case. As mentioned earlier the
conceptual difficulty is an a {\em par} with the one involved in the
Isomonodromic deformation system studied in Thm. \ref{cool} but we
need to set up a good deal of notation.

Let $\mathcal D$ be an effective divisor on $\C P^1$
\be
\mathcal D = r_\infty \infty + \sum_{\gamma\in \C} (r_\gamma+1)  \gamma\ ,\ \ r_\infty \geq 1, \ r_\gamma\geq 0.
\ee
It is understood that only a finite number of points $\gamma$ appear in the above sum.
We now consider an arbitrary rational connection of the form
\bea
A :=  \sum_{\gamma\in supp( \mathcal D)} A_\gamma(x) \\
A_\gamma(x) := \sum_{J=0}^{r_\gamma} \frac {A_{\gamma,J}}{(x-\gamma)^{J+1}}\\
A_\infty(x) := \sum_{J=1}^{r_\infty} x^{J-1} A_{\infty,J} \label{eq:generalA}
\eea
where the leading terms of the singularity at each irregular
singularity ($r_\gamma > 0$) may have an arbitrary Jordan canonical form
provided that the Lidskii pseudo-eigenvalues of the second leading
term subordinated to the Jordan form of the leading term do not vanish
and are distinct. No restriction is assumed on the residues at the Fuchsian singularities.
Note also that we are assuming that there is at least one irregular singularity and we have placed it at $\infty$.
%
%
%
%
%
%
\subsection{Generalized monodromy data for resonant singularities}
We now proceed along the lines of \cite{sibuya} to define the
generalized monodromy data for our system of linear ODE
\begin{eqnarray}\label{eq:iso}
{d\Psi\over{dx}}=A\Psi
\end{eqnarray}

Let $s_\gamma$ be the number of blocks in the Jordan form of
$A_{\gamma,r_\gamma}$ and $n_i^\gamma$ be the dimensions of the blocks
for a non-Fuchsian singularity ($r_\gamma\geq 1$).  Denote by
$R_\gamma$ be the diagonal matrix
\begin{eqnarray*}
R_\gamma={\rm diag}\left(\frac {t_{0,1}^\gamma}{n_1^\gamma},{{t_{0,1}^\gamma-1}\over
n_1^\gamma},\ldots,{{t_{0,1}^\gamma-n_1^\gamma+1}\over
n_1^\gamma},\ldots,\frac{t_{0,s}^\gamma}{n_s^\gamma},{{t_{0,s}^\gamma-1}\over
n_s^\gamma},\ldots,{{t_{0,s}^\gamma-n_s^\gamma+1}\over n_{s_\gamma}^\gamma}\right)
\end{eqnarray*}
where $t_{0,k}^c$ are scalars.

We can now use theorem \ref{main} and the results in Section
\ref{se:fuch} to construct formal solutions of (\ref{eq:iso}) near
each pole.

\begin{proposition}\label{pro:formalsol}
Let $A$ be given by (\ref{eq:generalA}) such that at each pole
$\gamma\in\mathcal D$, $A_{\gamma,r_\gamma}$ has the Jordan normal form
\begin{eqnarray*}
[A_{\gamma,r_\gamma}] = (\lambda_{\gamma,1}^{n^\gamma_1})\cdots
(\lambda_{\gamma,s}^{n^{\gamma}_s})
\end{eqnarray*}
and that the genericity condition in definition \ref{de:generic}
is satisfied at each pole if $r_\gamma\neq 0$. Then for each pole such
that $r_\gamma\neq 0$, there exists a unique formal series $Y^\gamma$
\begin{eqnarray}\label{eq:Yseries}
Y^\gamma(x-\gamma) &=& Y_0^\gamma + \sum_{j=1}^{\infty} (x-\gamma)^{j} Y_j^\gamma\ ,\ \
\det(Y_0^\gamma)\neq 0\
\end{eqnarray}
such that
\begin{eqnarray}\label{eq:formalsol}
\Psi^\gamma=Y^\gamma\exp(Q_\gamma(x-\gamma))(x-\gamma)^{R_\gamma}
\end{eqnarray}
is a formal solution to the ODE (\ref{eq:iso}), where $Q_\gamma(x-\gamma)$ is the matrix
\begin{eqnarray}\label{eq:expfactor}
Q_\gamma=\pmatrix{\ds \sum_{j=1}^{rn_1^\gamma}t_{j,1}^\gamma{{\mathcal H_{\gamma,1}^j}\over
j}&0&\ldots &0 \cr
           0&\ds \sum_{j=1}^{r \,n_2^\gamma}t_{j,2}^\gamma{{\mathcal H_{\gamma,2}^j}\over j}&\ldots &0\cr
           \vdots&\vdots&\ddots&\vdots\cr
           0&\ldots &0&\ds\sum_{j=1}^{rn_s^\gamma}t_{j,s}^\gamma{{\mathcal H_{\gamma,s}^j}\over j}\cr}\
\end{eqnarray}
where $\mathcal H_{\gamma,i}$ are $n_{i}^\gamma\times n_{i}^\gamma$ matrices defined as in
(\ref{shear}) and $t_{j,k}^\gamma$ are scalars.

When $r_\gamma=0$, there exist a solution of the form
\be\label{eq:formalFuch}
\Psi^\gamma = Y^\gamma (x-\gamma)^{G_\gamma}(x-\gamma)^{J^\gamma}\
. \
\ee
where $G_\gamma$ is a matrix determined by the integral
differences between the eigenvalues of $A_{\gamma,r_\gamma}$ as in section
\ref{se:fuch}, $Y^\gamma$ is a convergent series as in (\ref{eq:Yseries}) and $J^\gamma$ is a constant
Jordan matrix.
\end{proposition}
The uniqueness of $Y^\gamma$ can be seen by first making a gauge
transformation of $A$ such that $A_{\gamma,r_\gamma}$ is in Jordan canonical
form.

We can now use this formal solution to define the monodromic data
of the ODE (\ref{eq:iso}). We have the following
\begin{theorem}\label{thm:mono}
There exists a finite number of asymptotic sectors $\mathcal S_i^\gamma$ with
vertex at $x=\gamma$, such that $\left\{\mathcal S_i^\gamma\right\}$ is a
covering at $x=\gamma$. Moreover, there exist fundamental solutions
$\Psi_i^{(\gamma)}$ of the linear ODE
\begin{eqnarray*}
{d\Psi\over{dx}}=A\Psi
\end{eqnarray*}
asymptotic to the formal series solution of the form (\ref{eq:formalsol}) or
(\ref{eq:formalFuch}) near $x=\gamma$.
\end{theorem}
The proof of this result can be found in \cite{sibuya} or
\cite{wasow}.
Such solutions $\Psi_i^\gamma$ are analytic outside of the divisor
$\mathcal D$ and on the universal cover of $\C P^1\setminus \mathcal
D$.

In particular, we can now cover the punctured Riemann sphere
$\mathbb{CP}^1/\mathcal D$ with sectors $\mathcal S_j^\gamma$,
$\gamma\in\mathcal D$ and define the stokes data by using the
asymptotic form (\ref{eq:formalsol}) and the sectors $\mathcal
S_j^c$.

We have the following
\begin{definition}\label{de:stokes}
Let $\left\{\mathcal S_i^\gamma\right\}$ be a covering of the punctured Riemann
sphere $\mathbb{CP}^1/\mathcal D$ that satisfies the conditions in
theorem \ref{thm:mono} and let $\Psi_i^\gamma$ be the fundamental
solution  that is asymptotic to the formal
solution of the form (\ref{eq:formalsol}) in $\mathcal S_i^\gamma$. Let
\begin{eqnarray*}
C_{kl}^{\alpha\beta}=\left(\Psi_k^{\alpha}\right)^{-1}\Psi_l^{\beta}\ .
\end{eqnarray*}
 We shall call $\left\{\mathcal S_i^\gamma,C_{kl}^{\alpha\beta}\right\}$ a Stokes
phenomenon of the  ODE (\ref{eq:iso}). This, together
with the set of variables $t_{0,j}^c$ at each pole $x=\gamma$, $r_\gamma\neq
0$, and the $J^\gamma$ for $r_\gamma=0$, is called the monodromic data of
the ODE (\ref{eq:iso}).
\end{definition}
Here the matrices $C_{kl}^{\alpha\beta}$ contain the usual definition of Stokes' matrices and connection and monodromy on the same basis (and redundantly).
\subsection{Isomonodromic deformations}
In view of Thm. \ref{cool}, at each singularity $\gamma\in \mathcal D$ we
  can find a formal nonsingular gauge $Y_\gamma$ that gauges $A(x)$ to the
  localized version of the bare form $D_{(\gamma)}(x)$ advocated
  Prop. \ref{lemma2finite}.
  The bare connection/deformation at $x=\gamma$ is now the direct sum of as
  many bare systems of the form given in Prop. \ref{lemmabarefinite}
  as the number of blocks in which the system has been decomposed. All
  these
 bare deformation/differential equations are then dressed by the same dressing matrix $Y^\gamma$.

 In order to be more explicit  consider an irregular
 singularity $x=\gamma$. Suppose the Jordan form of the leading coefficient
 at $x=\gamma$ is as in Prop. \ref{pro:formalsol}.
In symbolic notation we denote by ${\bf t}_\gamma$ the times ``attached''
 to the pole at $x=\gamma$. The number of them is
\be
\#\{{\bf t}_\gamma\} = rn\ ,
\ee
namely the same as if the system were nonresonant.

The following theorem allows us to consider the direct sum of many
``bare'' Pfaffian systems of the form in Lemmas \ref{lemmabare},
\ref{lemmabarefinite} and obtain a solution of another Pfaffian
systems after a ``dressing'' using a formal gauge equivalence.
\begin{theorem}\label{thm:bridge}
Suppose $A$ is a matrix-valued function given by
(\ref{eq:generalA}) and $Y^\gamma$ is a formal series holomorphic and
invertible at $\gamma$
\begin{eqnarray*}
Y^\gamma(x-\gamma) &=& Y_0^\gamma + \sum_{j=1}^{\infty} (x-\gamma)^{j} Y_j^\gamma\ ,\ \
\det(Y_0)\neq 0\ \\ Y^{\infty}(x) &=& Y_0^{\infty} +
\sum_{j=1}^{\infty} x^{-j} Y_j^{\infty}
\end{eqnarray*}
such that
\begin{eqnarray*}
B^\gamma(x)= \left(Y^\gamma\right)^{-1}AY^\gamma -
\left(Y^\gamma\right)^{-1}\left(Y^\gamma\right)'\ ,
\end{eqnarray*}
is a formal power series in $(x-\gamma)$ in block diagonal form
according to the block decomposition of $A_{\gamma,r_\gamma}$, and similar
expression for $\infty$.

Let $\mathcal T^\gamma(x)$ ($\mathcal T^\infty(x)$)  be matrix-valued 1-forms
polynomial in $(x-\gamma)^{-1}$ ($x$, respectively) such that
\bea
\le[\pa_x - B^\gamma(x),{\rm d} - \mathcal T^\gamma(x)\ri]&=& {\mathcal
O}(x-\gamma) \\
\le[\pa_x - B^{\infty}(x),{\rm d} - \mathcal T^{\infty}(x)\ri]&=&
{\mathcal O}(x^{-1})\ .
 \eea
Then the Pffaffian system
\bea
\le[\pa_x - A(x)\ri]\Psi = 0\\
\le[{\rm d} - \mathcal M(x)\ri]\Psi=0 \eea is Fr\"obenius
compatible, where $\mathcal M(x)$ is the matrix-valued rational
1-form such that the singular part of $\mathcal M$ near each pole
is given by \bea \mathcal M(x)=\le(Y^\gamma \mathcal
T^\gamma(x-\gamma)\le(Y^\gamma\ri)^{-1}\ri)_{pp}+{\mathcal
O}(1),\quad x\rightarrow \gamma \neq \infty \eea where $X_{pp}$
denotes  the principal part of $X$ at $x=\gamma$. Near $x=\infty$,
$\mathcal M$ is given by \bea \mathcal M(x)=\le(Y^{\infty}\mathcal
T^{\infty}(x)\le(Y^{\infty}\ri)^{-1}+{\rm
d}Y^{\infty}\le(Y^{\infty}\ri)^{-1}\ri)_{+}+{\mathcal
O}(x^{-1}),\quad x\rightarrow\infty \eea where $X_+$ is the
polynomial part of $X$ and $Y_0^{\infty}$ is some analytic
function in the deformation parameters.
\end{theorem}
Proof. Let $\mathcal M (x)=\sum \mathcal M_{\nu}(x)dt_{\nu}$: here the
label $\nu$ is a generic label used for the collection of all times at
all singularities. According to this definition each $\mathcal M_\nu$
is singular only at one of the poles $\gamma\in(\mathcal D)$. Let $\nu$ be
the pole of $ \mathcal M_{\nu}(x)$. A simple calculation shows that
\begin{eqnarray*}
\le[\pa_{\nu},\pa_x\ri]Y^{\nu}\le((\hat Y)^{\nu}\ri)^{-1}=\pa_{\nu}
A -\pa_{x} \mathcal M_{\nu}(x)+ [A, \mathcal M_{\nu}(x)] -\le( \pa_{\nu} B^{\nu}
-\pa_{x}T_{\nu}^{\nu} + [B^\nu,\mathcal T_{\nu}^{\nu}(x)]\ri)
\end{eqnarray*}
where $M_{\nu}(x)=\le({Y}^{\nu}\mathcal
T_{\nu}\le({Y}^{\nu}\ri)^{-1}\ri)_{pp}$.

By using similar argument as in the proof of theorem \ref{cool},
we see that
\begin{eqnarray*}
\le[\pa_x - A(x),\pa_{\nu} -   \mathcal M_{\nu}(x)\ri]&=&{\mathcal
O}(x-\nu),\quad x\rightarrow \nu \\
\le[\pa_x - A(x),\pa_{\nu} -  \mathcal M_{\nu}(x)\ri]&=&{\mathcal
O}(x^{-1}),\quad x\rightarrow\infty
\end{eqnarray*}
Therefore, by Liouville's theorem, we have
\begin{eqnarray*}
\le[\pa_x - A(x),\pa_{\nu} - \mathcal  M_{\nu}(x)\ri]=0
\end{eqnarray*}
Similarly, we have
\begin{eqnarray*}
\le[\pa_{\nu} - \mathcal M_{\nu}(x),\pa_{\mu} - \mathcal  M_{\mu}(x)\ri]&=&{\mathcal
O}(x-\nu),\quad x\rightarrow \nu=\mu \\
\le[\pa_{\nu} - \mathcal  M_{\nu}(x),\pa_{\mu} - \mathcal  M_{\mu}(x)\ri]&=&{\mathcal
O}(x^{-1}),\quad x\rightarrow \nu=\mu=\infty
\end{eqnarray*}
when the pole of $ \mathcal M_{\nu}(x)$ and $ \mathcal M_{\mu}(x)$ are the same point.

Now consider the case when the poles of $ \mathcal M_{\nu}(x)$ and
$ \mathcal M_{\mu}(x)$ are not the same. Then near
$x=\nu$, we have
\begin{eqnarray*}
\pa_{\nu}Y^{\nu}&=& \mathcal M_{\nu}(x)Y^{\nu}+Y^{\nu} \mathcal T_{\nu}^{\nu} \\
\pa_{\mu}Y^{\nu}&=& \mathcal M_{\mu}(x)Y^{\nu}
\end{eqnarray*}
We also have similar equations with $\nu$ replaced by $\mu$.
From this we have
\begin{eqnarray*}
\le[\pa_{\nu},\pa_{\mu}\ri]Y^{\nu}\le(Y^{\nu}\ri)^{-1}=\pa_{\nu}
 \mathcal M_{\nu}(x) -\pa_{\mu} \mathcal M_{\nu}(x)+ [ \mathcal M_{\mu}(x), \mathcal M_{\nu}(x)]
\end{eqnarray*}
near $x=\nu$ and similar equation with $\mu$ replacing
$\nu$. By using similar argument as before, we see that
\begin{eqnarray*}
\pa_{\nu}  \mathcal M_{\nu}(x) -\pa_{\mu} \mathcal M_{\nu}(x)+
[ \mathcal M_{\mu}(x), \mathcal M_{\nu}(x)]=0
\end{eqnarray*}
this concludes the proof of the theorem. Q.E.D.

We can now consider isomonodromic deformations of the ODE
\begin{eqnarray*}
{{d\Psi}\over{dx}}=A\Psi
\end{eqnarray*}
Let $\le\{\mathcal S_k^{\alpha},\Psi_k^{\alpha}\ri\}$ be a
covering and fundamental solutions as in definition
\ref{de:stokes} and let
$\left\{C_{kl}^{\alpha\beta},J^\gamma,t_{0,k}^\gamma\right\}$ be the
monodromic data associated to it. The isomonodromic problem is the
following. Given $\le\{\mathcal
S_k^{\alpha},\Psi_k^{\alpha}\ri\}$, how should one deform the
solutions $\Psi_k^{\alpha}$ such that the monodromic data
$\left\{C_{kl}^{\alpha\beta},J^\gamma,t_{0,k}^\gamma\right\}$ is fixed?

The isomonodromic problem is only defined when a covering and
solutions $\le\{\mathcal S_k^{\alpha},\Psi_k^{\alpha}\ri\}$ is
chosen and should be thought of as deformations of the fundamental
solutions. For if the monodromic data of $\le\{\mathcal
S_k^{\alpha},\Psi_k^{\alpha}\ri\}$ is fixed under deformations,
then by multiplying $\Psi_k^{\alpha}$ to the right by matrices
$C_k^{\alpha}$ that depend on the deformation parameters $t$, one
obtains a pair $\le\{\mathcal
S_k^{\alpha},\Psi_k^{\alpha}C_k^{\alpha}\ri\}$ in which the
monodromic data is no longer fixed.

We can now prove the following theorem, which classifies the
monodromy preserving deformations when the genericity
condition in definition \ref{de:generic} is satisfied.

In the following we denote the matrix-differential one form of the bare deformations as
\be
\mathcal T^{\gamma,bare}(x) = {\rm diag}\le(\mathcal C_1^\gamma{\rm d}\gamma + \sum_{J=1}^{rn_1^\gamma} T_{1}^{\gamma,bare}{\rm d}t_{J,1}^\gamma ,\dots,  \mathcal C_{s_\gamma} ^\gamma{\rm d}\gamma+\sum_{J=1}^{rn_{s_\gamma}^{\gamma}}T_{s_\gamma}^{\gamma,bare}{\rm d}t_{J,s_\gamma}^{\gamma}\ri)
\label{genbaredef}
\ee
where the notation is as in Props. \ref{lemmabare}, \ref{lemmabarefinite}.

\begin{theorem}\label{thm:isogen}
Let $A$ be a rational matrix-valued function given by
(\ref{eq:generalA}) and let $\mathcal D_0$ be the divisor of poles
such that $r_\gamma=0$ for $\gamma \in\mathcal D_0$ and $D_1$ be the divisor
of higher order poles. Suppose the genericity condition in
definition \ref{de:generic} is satisfied at each pole
$\gamma \in\mathcal D_1$ and that $\infty\in\mathcal D_1$.

Let $\le\{\mathcal S_k^{\alpha},\Psi_k^{\alpha}\ri\}$ be a
covering and fundamental solutions that satisfies the conditions
in Def. \ref{de:stokes}. Then the monodromic data of the ODE
(\ref{eq:iso}) defined by $\le\{\mathcal
S_k^{\alpha},\Psi_k^{\alpha}\ri\}$ are preserved if and only if
$\Psi_k^{\alpha}$ satisfies the differential equations
\begin{eqnarray}\label{eq:deform}
{\rm d}\Psi_k^{\alpha}=\mathcal T \Psi_k^{\alpha}
\end{eqnarray}
where $\mathcal T$ is the 1-form
\begin{eqnarray*}
\mathcal T&=&\left(Y^{\infty}\mathcal T^{\infty,bare}(Y^{\infty})^{-1}+{\rm
d}Y_0^{\infty}\le(Y_0^{\infty}\ri)^{-1}\right)_{+} +
\sum_{\gamma\in\mathcal D_1}\left(Y^\gamma \mathcal T^{\gamma,bare}(Y^\gamma)^{-1}\right)_{pp}+\\
&+&\sum_{\gamma\in\mathcal D_0}A_{\gamma,0}(x-\gamma)^{-1}{\rm
d}\gamma+\le(Y^\gamma(x-\gamma)^{G_\gamma}{\rm
d}H^\gamma(H^\gamma)^{-1}(x-\gamma)^{-G_\gamma}\le(Y^\gamma\ri)^{-1}\ri)_{pp}
\end{eqnarray*}
where $T^{\gamma,bare}$ is defined in (\ref{genbaredef})  and $Y^\gamma$, $G_\gamma$,  and $Y^{\infty}$
are given in proposition \ref{pro:formalsol}, while $H^\gamma$ is a matrix in the
centralizer of $J_\gamma$ which is constant in $x$ and depends analytically on the deformation parameters. We use $X_{pp}$ to denote the
principal part of $X$ around the corresponding pole and $X_+$ to
denote the polynomial part of $X$. The exterior derivative $d$ in
the above expressions denotes derivatives with respect to the $t$
in proposition \ref{pro:formalsol} and the position of the poles
$\gamma$, but with d$t_{0,k}^\gamma=0$. The dependence of $Y_0^{\infty}$ and
$H^\gamma$ on the parameters, that is, ${\rm d}Y_0^{\infty}$ and ${\rm
d}H^\gamma$, can be arbitrary, provided that $H^\gamma$ remains in the centralizer of $J_\gamma$.
\end{theorem}
{\bf Proof}. The proof is essentially the one in \cite{JMU}.
By using similar argument as in the proof of theorem
\ref{thm:bridge}, we see that in the case of $r_\gamma=0$, we have
\begin{eqnarray*}
\le[\pa_{x},\pa_{\gamma}\ri]Y^\gamma\le(Y^\gamma\ri)^{-1}+Y^\gamma\le[\pa_{x},\pa_{\gamma}\ri]\le((x-
\gamma)^{G_\gamma}H^\gamma
\ri)\le(Y^\gamma(x-\gamma)^{G_\gamma}H^\gamma\ri)^{-1}
=\pa_{x}
\mathcal T_\gamma(x) -\pa_{\gamma}A(x)+ [ A(x), \mathcal T_{\gamma}(x)]
\end{eqnarray*}
near $x=\gamma$ where $\mathcal T_\gamma$ is the coefficient of ${\rm d}\gamma$ in $\mathcal T$. The left hand side is a positive series in $x-\gamma$ since the
second term is a well-defined function of $x$ and the deformation parameters,  and $Y^\gamma$ is a positive series in $x-\gamma$. By considering the singular
behavior of the right hand side near each pole and then apply
Liouville's theorem, one sees that the right hand side must be
zero. Similarly, one can show the commutativity between
the parameters $t$, $\gamma$ when one of the parameters involved is the
position of a simple pole.

For the cases that do not involve derivatives of the position of
a simple pole, we can apply proposition \ref{lemmabarefinite} and
theorem \ref{thm:bridge}. We see that the Pffafian system
\begin{eqnarray}\label{eq:com}
\le\{ \begin{array}{l}
\le[\pa_x - A(x)\ri]\Psi = 0\\[5pt]
\le[{\rm d} - \mathcal T(x)\ri]\Psi=0
\end{array}
\ri .
\end{eqnarray}
is integrable.

Let $\Psi_k^{\alpha}$ and $\Psi_l^{\beta}$ be solutions of
(\ref{eq:com}) as in Def. \ref{de:stokes}, where
$\alpha$, $\beta\in\mathcal D$ and may not be the same point.
Since $\Psi_k^{\alpha}$ and $\Psi_l^{\beta}$ solve the same
equations (\ref{eq:com}), we have
\begin{eqnarray*}
{\rm d}\Psi_k^{\alpha}\le(\Psi_k^{\alpha}\ri)^{-1}={\rm
d}\Psi_l^{\beta}\le(\Psi_k^{\beta}\ri)^{-1}={\rm
d}\Psi_k^{\alpha}\le(\Psi_k^{\alpha}\ri)^{-1}+\Psi_k^{\alpha}{\rm
d}C_{kl}^{\alpha\beta}\le(C_{kl}^{\alpha\beta}\ri)^{-1}\le(\Psi_k^{\alpha}\ri)^{-1}
\end{eqnarray*}
hence ${\rm d}C_{kl}^{\alpha\beta}\le(C_{kl}^{\alpha\beta}\ri)^{-1}=0$ and the
monodromic data is preserved.

Conversely, let $\Psi_k^{\alpha}$ be deformed  in such a way that the monodromic data
defined by $\le\{\mathcal S_k^{\alpha},\Psi_k^{\alpha}\ri\}$ is
preserved. We see that
\begin{eqnarray*}
{\rm d}\Psi_k^{\alpha}\le(\Psi_k^{\alpha}\ri)^{-1}={\rm
d}\Psi_l^{\beta}\le(\Psi_k^{\beta}\ri)^{-1}
\end{eqnarray*}
for $\alpha$, $\beta\in\mathcal D$, where $\alpha$ and $\beta$ may
not be the same point. Therefore ${\rm
d}\Psi_k^{\alpha}\le(\Psi_k^{\alpha}\ri)^{-1}$ is a globally
defined meromorphic 1-form
\begin{eqnarray}\label{eq:com1}
 \mathcal M(x):= {\rm d}\Psi_k^{\alpha} \le({\Psi_k^{\alpha}}\ri)^{-1}\ .
\end{eqnarray}
 Its asymptotic behavior near each pole
are given by (note that ${\rm d} Q_\gamma = \mathcal T^{\gamma,bare}$)
\begin{eqnarray}\label{eq:singpart}
{\rm d}\Psi_k^{\alpha}\le(\Psi_k^{\alpha}\ri)^{-1}&=&{\rm
d}\Psi_l^{\beta}\le(\Psi_l^{\beta}\ri)^{-1}\sim {\rm
d}Y^\gamma\le(Y^\gamma\ri)^{-1}+Y^\gamma{\rm
d}{Q_\gamma}\le(Y^\gamma\ri)^{-1}\nonumber \\
&=&\le(Y^\gamma{\rm d}{Q_\gamma}\le(Y^\gamma\ri)^{-1}\ri)_{pp}+{\mathcal
O}(1)\qquad \qquad x\rightarrow \gamma,\quad r_\gamma>0 \\
{\rm
d}\Psi_k^{\alpha}\le(\Psi_k^{\alpha}\ri)^{-1}&=&A_{\gamma,0}(x-\gamma)^{-1}{\rm
d}\gamma+\le(Y^\gamma(x-\gamma)^{G_\gamma}{\rm
d}H^\gamma(H^\gamma)^{-1}(x-\gamma)^{-G_\gamma}\le(Y^\gamma\ri)^{-1}\ri)_{pp} + \mathcal
O(1)\nonumber \\
&& x\rightarrow \gamma,\quad r_\gamma=0
\end{eqnarray}
for all $\alpha$, $\beta$ and $\gamma\in\mathcal D$, where $X\sim W$
means that $X$ is asymptotic to $W$ near the corresponding point.
One can also write down a similar equation for
$x\rightarrow\infty$.

The equations (\ref{eq:singpart}) determines the 1-form $\mathcal
M$ up to the addition of a constant in $x$.

To determine this constant, we can look at the behavior of
$\mathcal M$ near $x=\infty$. This is given by
\begin{eqnarray*}
\le({\rm d}Y^{\infty}\le(Y^{\infty}\ri)^{-1}+Y^{\infty}{\rm
d}Q_{\infty}\le(Y^{\infty}\ri)^{-1}\ri)_+\quad x\rightarrow \infty
\end{eqnarray*}
We see that $\mathcal M$ has the form of $\mathcal T$ in the
theorem. Q.E.D.\par \vskip 5pt
\br\label{RHrem} The essence of Thm. \ref{thm:isogen} is that the
dependence of $Y^{\infty}$ and $H^\gamma$ can be regarded as gauge
arbitrariness, and fixing it yields a consistent Pfaffian system.

Suppose that we have an initial value problem for the Pfaffian system
(\ref{eq:com}) $A^{(0)}(x)$ and that we consider two evolutions, one
in which $H^\gamma$ are constants in the parameters and one in which
they are preassigned arbitrary analytic functions with values in the
prescribed centralizers. Let us denote by $\Psi$ and $\widetilde \Psi$
the two kernel solutions of the Pfaffian system.
Define then the function $G(z,{\bf t})$ (here  ${\bf t}$ denotes
collectively all the isomonodromic deformation parameters) as follows
\be
G(x,{\bf t}) = \widetilde \Psi_k^\alpha(x)(\Psi_{k}^\alpha)^{-1}(x)\ .
\ee
Since both Stokes' phenomena are the same, this implies that $G(z)$ is
a single-valued analytic invertible function on the punctured domain
$\C \mathbb P^1\setminus \mathcal D$.
 Since the essential singularities of $\Psi$ and $\widetilde \Psi$
 have the same asymptotic expansion in the same sectors of the
 irregular singularities, this implies that $G(x)$ has no singularity
 there. The only possible singularities are poles at the Fuchsian
 singularities. The same considerations used in  Section
 (\ref{se:fuch}) show that in general $G(x)$ has poles of order equal
 to the resonance index at the given Fuchsian singularity (in case of
 nonresonant singularity $G(x)$ must be analytic there). This shows
 that the RH problem for $G$ discussed in Sect. \ref{se:fuch} admits a
 solution for arbitrary choice of the group elements $H^\gamma$ in the
 centralizer of the local monodromy matrices.
 In other words the dependence of the  $H^\gamma$'s on the parameters
 ${\bf t}$ is ``pure gauge'' and can be completely gauged away by
 means of a {\bf rational} gauge equivalence $G(x,{\bf t})$ which does
 not alter the singularity structure of $A(x)$ .
\er

In order to fix part of this arbitrariness we can assume that
the coefficient $A_{\infty,r_\infty}$ is in the gauge-fixed form
(\ref{gaugefixing}) within each block: this fixes an overall gauge
$Y^\infty$ uniquely (up
to the centralizer of $A_{\infty,r_\infty}$) for our connection. At
this point the procedure of Thm. \ref{main} for the singularity at
$\infty$ produces a unique formal gauge $Y_\infty$ in which the
leading coefficient is necessarily in the centralizer of
$A_{\infty,r_\infty}$; the entries of this leading coefficient will be
constants of the motions exactly as in the previous guide-example, and
hence could be set to zero in the sense that this would be a
consistent reduction of the problem (but of no practical advantage).

\bc[Normalized isomonodromic deformations]
\label{cor:normal}
Let $A(x)$ be as in Thm. \ref{thm:isogen}. Then the one form obtained
from eq. (\ref{eq:deform}) by setting ${\rm d}Y^{\infty} \equiv
0\equiv {\rm d}H^\gamma$ defines an integrable Pfaffian
system.
\ec

%
%
\section{Isomonodromic Tau function}
The authors of \cite{JMU} gave a definition of tau function in terms
of a suitable closed differential on the space of times. The
definition was expressed in terms of (formal) residues of  the formal
local gauge (which we denote $Y^\gamma$) around each singularity of the
connection. It is shown in \cite{BHHP} that that definition is
equivalent to one which manifests the spectral nature of the tau
function. In fact the definition in \cite{BHHP} is much more
convenient and easily generalizable to the present context

We first need to fix the gauge arbitrariness of our deformation system
by requiring that the deformations occur according to the integrable
Pfaffian system specified in Corollary \ref{cor:normal}.

Let us set up the notation: we consider the {\bf spectral curve}
(characteristic polynomial) of the connection, namely the set of
$\C^2$
\be
E(x,y):= \det(y-A(x)) = \sum_{j=0}^{n} y^j P_{n-j}[x;A] = 0\ .
\ee
We think of this curve $\Sigma$ in $\mathbb P^1\times \mathbb P^1$ and
we regard the two functions $x,y:\Sigma\to \C P^1$ as meromorphic
functions on the curve, or projections.
The $x$ projection  $x:\Sigma\to \mathbb C P^1$ has ramification
points at all the $\gamma \in \mathcal D_1$ (irregular singularities) and
at $\infty$, together with other ramification point whose position is
not known a priori and in general depends on the "times".

If we perform an appropriate desingularization of the curve above each
pole $x=\gamma$ of the divisor $\mathcal D_1$ (including $\infty$)  we obtain a branching structure in
which the $n$ sheets of the projection $x:\Sigma\to \C P^1$ are glued
precisely according to the dimensions of the Jordan cells of the
leading coefficient of the connection and cyclically permuted by a
small loop of $x$ around $\gamma$.
For example, suppose  the Jordan form of $A_{(\gamma),r}$ is (we set
$r_\gamma=r$ for brevity since the discussion is local anyway)
\be
[A_{(\gamma),r}] =({\lambda_1}^{5})^2 {\lambda_2}^3 \lambda_3 (\lambda_4)^{2}
\ee
This means that there are $2$ Jordan blocks of size $n_1=5=n_2$ and
eigenvalue $\lambda_1$, one Jordan block of size $3$ and eigenvalue
$\lambda_2$ etc. (this would correspond to $n = 5\times 2 + 3 + 1+
1\times 2 = 16$). Under our assumptions for the Lidskii matrix of the
subleading term, these eigenvalues split into {\em distinct} cyclic $k-$tuplets where
$k$ is the size of the block and the number of $k-$tuplets equals the
number of blocks of that size. In the example we will have two groups
of five-plets $y_{1,j}, y_{2,j}$, $j=1\dots n_1=n_2$ that have the common
asymptotics
\be
y_{a,j}\simeq \lambda_1 (x-\gamma)^{-1-r} (1+ o(1)).
\ee
Both groups admit a Puiseux series expansion in powers of $\xi:=(x-\gamma)^{\frac
  1 {n_1}}$
and each group forms a cyclic $n_1$-tuplet
\bea
y_{a,j} = \frac{\lambda_1 }{\xi^{n_a(r+1)}} \le(1 + \sum_{k=1}^\infty
c_{a,k} \xi^k\ri)\\
y_{a,j}({\rm e}^{2i\pi/n_a} \xi) = y_{a,j+1} (\xi).
\eea
The two groups are distinct in the sense that the first coefficients
of the expansion in the brackets differ from each other $c_{a,1}\neq
c_{b,1}$, $a\neq b$.
The first $r\cdot n_a = n_a\,r_c$ coefficients are actually the isomonodromic
 times of our problem, or more precisely (carrying on with this example)
\be
y_{a,1}  = \frac {1}{n_a\xi^{n_a(r_c+1)}}\le( t_{a,n_a r} +
  \sum_{K=1}^{n_1r-1} t_{a,n_ar-K} \xi^{K} + \dots \ri) =\sum_{J=1}^{n_ar}
  \frac {t_{a,J}}{n_a\xi^{n_a+J}} +  \frac {\widetilde t_{a,0}}{n_a \xi^{n_a}} +
 \sum_{K=1}^{\infty}
  \frac  {K\,H_{a,K}}{n_a} \xi^{K} \ .
\ee This means that we can ``extract'' the isomonodromic times (as
well as other parameters of formal monodromy) with the residue
\bea t_{a,J} &=& \res{x=\gamma}  \le(\sum_{\sigma=0}^{n_a-1}
(\omega_a)^{J\sigma}y_{a,\sigma+1}\,n_a\xi^{\frac J {n_a}}\ri){\rm
d}x = \res{\zeta_\gamma^1} y (x-\gamma)^{\frac J{n_a}} {\rm d}x  \
,\nonumber \\ J&=&1,\dots, n_ar\ ,\ \
\omega_a:= {\rm e}^{2i\pi/n_a} \nonumber \\
\widetilde t_{a,0}&=& t_0 - \frac {n_a-1}2 = \res{x=\gamma }
\le(\sum_{\sigma=0}^{n_a-1} y_{a,\sigma+1}\ri){\rm d}x = \res{\zeta_\gamma^a}
y {\rm d}x
\eea
Note that the fractional power is in fact a well-defined local
coordinate on the (desingularization of the) spectral curve. The
notation is that $\zeta_\gamma^a$ is the point on the spectral curve
that projects to $\gamma=x(\zeta_\gamma^a)$ and $a$ distinguishes the
different (in this case $a=1,2$) points projecting to the same
$\gamma$ (and corresponds to different Jordan cells of the same
dimension, in this example $n_a=5$, $a=1,2$).

We now define the isomonodromic tau function  by the following
equations \bea \frac\pa{\pa t_{a,K}} \ln \tau :&=& H_{a,K} = \frac
1{K} \res{x=\gamma}
\le(\sum_{\sigma=0}^{n_a-1} y_{a,\sigma+1} (\omega_a)^{J\sigma}(x-\gamma)^{K/n}\ri) {\rm d}x\\
\frac \pa{\pa \gamma} \ln \tau :&=& \frac 1 2 \res{x=\gamma} \tr
A^2(x){\rm d}x\label{movepole} \eea Quite clearly the above
definition should be repeated for all times and all singularities
of our connection, where for the Fuchsian singularities the only
pertinent equation is (\ref{movepole}): then we should {\em
  prove} that the definition is well-posed, namely that the
differential defined by these equation is closed.

In fact -as it turns out- the proof of this fact is no different from
\cite{BHHP} and the previous simple case of polynomial connection
with single Jordan leading block, needing just some modifications in
the notation.

It is based mainly on the following
\bl
\label{bidiff}
Let $A(x)$ be a rational matrix and let $\widetilde M$ denote the
classical adjoint of a matrix $M$. Let $y(x)$ be the (multivalued)
eigenvalue of $A(x)$. Define the following expression
\be
\Omega := \frac {\tr \le(\widetilde{(y-A)}(x)\, \widetilde
      {(y'-A)} (x')\ri)}{\tr \le(\widetilde{(y-A)}(x)\ri)
  \,\tr\le(\widetilde{(y'-A)}(x')  \ri)} \frac{{\rm d}x{\rm
    d}x'}{(x-x')^2}\ ,
\ee
where $y(x)$ and $y'(x)$ are two determination of the eigenvalues above the point $x$.
Then $\Omega$ is a well defined bidifferential on the spectral curve,
with a double pole without residue on the diagonal away from the singularities of $A$ and from the branch-points of $y(x)$.
\el
Note that the lemma states that -in spite of the $(x-x')^2$
denominator, there are no singularities unless $y(x')\to y'(x)$ as
$x'\to x$ (i.e. unless $y$ and $y'$ belong to the same sheet of the $x$-projection).

{\bf Proof}. This is essentially the same argument used in  the proof of Prop. (\ref{pro:tausimple}).
The only -a priori- singularities of  $\Omega$ are at the poles of $A$
and possibly at the branch-points of $y$ -which are of no interest to
us- {\bf and} on the points above $x=x'$. We must prove that if $y(x)$
and $y(x')$ are on different sheets then there is no pole and if they
belong to the same sheet then there is a double pole without residue.
Let $x$ and $x'$ be in a common neighborhood which does not include
any  branch-points; then we can distinguish the $r$ sheets of the spectral curve $y_1,\dots,y_r$. The expressions
\be
\Pi_a(x):= \frac {\widetilde {A-y_a}(x)}{\tr(\widetilde {A-y_a})(x)}\ ,
\ee
are then the rank-one spectral projectors on the (one-dimensional)
eigenspace and they are well defined in said neighborhood. As we
already remarked we have
\be
\tr(\Pi_a(x)\Pi_b(x')) = \delta_{ab} + \mathcal O ( (x-x')^2)
\ee
which proves the assertion. Q.E.D.\par\vskip 5pt
\br
In fact one may prove a stronger assertion (which is not important for
our purposes) that $\Omega$ is a well defined differential on the
spectral curve everywhere except the diagonal $(x,y)=(x',y')$.
\er
The bidifferential $\Omega$ appears naturally when computing the
closure of the differential of the tau function and the absence of
residue on the diagonal is the key feature that allows proving such
closure.

We now remark that --in the same way as in Section
\ref{taufunctionsimple}-- the formal gauge $Y^\gamma(x)
(x-\gamma)^{G_\gamma}W^\gamma$ near any irregular singularity
coincides with the eigenvector matrix up to high order. More
precisely, near $x=\gamma$ ($r_\gamma>0$) we have \bea A(x) =
(Y^{\gamma})^{-1} D^\gamma(x) Y^\gamma - (Y^{\gamma})^{-1} \frac
{{\rm d}}{{\rm d}x} Y^\gamma \eea where $D^\gamma(x)$ is the block
diagonal matrix (we momentarily suppress the explicit reference to
$\gamma$ from the labels for notational convenience) \bea D(x) &=&
\pmatrix{\ds \sum_{J=0}^{rn_1}t_{J,1}{{\mathcal H_{1}^J}\over
{n_1(x-\gamma)}}& & \cr &\ddots &\cr
             &&\ds\sum_{J=0}^{rn_s^\gamma}t_{J,s}{{\mathcal H_{s}^J}\over
             {n_s(x-\gamma)}}\cr} \!\!+\!\!
\pmatrix{\ds
\frac {G_1}{x-\gamma}  & &\cr
& \ddots &\cr
& & \ds\frac {G_s}{x-\gamma}
}=\\
&=& {\rm diag}(D_1(x),\dots,D_s(x))\label{diagbare}\\
&&G_j:= {\rm diag}\le(0,\frac 1 {n_j},\dots,\frac{n_j-1}{n_j}\ri)
\eea
Define now the block-diagonal matrices
\bea
&&G:={\rm diag} (G_1,\dots, G_s)\\
&&W:= {\rm diag}(W_1,\dots,W_2)\ ,\ \ \ W_j:=
    [\omega_j^{-(\ell-1)(k-1)}]_{\ell,k}\ ,\ \ \omega_j:= {\rm
      e}^{\frac {2i\pi}{n_j}}\ .
\eea
Then
\bea
A &=& Y x^{-G} W {\rm  diag}( \hat D_1(x),\dots,\hat D_s(x)) W^{-1}x^G Y^{-1} + Y'Y^{-1}\\
\hat D_j(x) :&=& \frac 1{n_j (x-\gamma)}\sum_{K=0}^{rn_j} t_K
(x-\gamma)^{\frac K {n_j}} {\Omega_j}^K - \frac {(x-\gamma)^{G_j}
  {W_j}^{-1}G_j W_j (x-\gamma)^{-G_j}}{n_j( x-\gamma)} \\
\Omega_j :&=& {\rm diag}
(1,\omega_j,{\omega_j}^{2},\dots,\omega_j^{n_j-1}) \eea As in
Section \ref{taufunctionsimple} this implies that $Z:= Y
(x-\gamma)^{-G}W$ coincides with an eigenvector matrix $P$ up to
order $(x-\gamma)^{1-r + \epsilon}$ where $\epsilon =
\min\le\{\frac 1{n_j}\ri\} $ and in turn this implies that the
deformation matrices are
\bea
J\mathcal T_{J,j} &\!\!\!\!=& \!\!\!\!\!\le( Y {\rm diag}(0,\dots,{\mathcal H_j}^{J},\dots,0)  Y^{-1} \ri)_{pp}\!\!\! =
\le( (x\!-\!\gamma)^{-\frac J {n_j}} Z {\rm diag}(0,\dots,W_j
   {\Omega_j}^J {W_j}^{-1},\dots,0) Z^{-1}\ri)_{pp}  \!\!\!= \nonumber
   \\
&\!\!\!\!=& \!\!\!\!\! \bigg( P {\rm diag}(0,\dots,W_j {\Omega_j}^J {W_j}^{-1},\dots,0) P^{-1}\bigg)_{pp}\label{specdef}\\
 j&=&1,\dots,s,\ \ J = 1,\dots, r\,n_j\ \\
\mathcal C &\!\!\!\!=& \!\!\!\!\! \le( Y \, D\, Y^{-1}
\ri)_{pp}\!\!\! = A_\gamma(x) \eea For the deformation of
parameters at $\infty$ the principal part is to be replaced by the
polynomial part and then the identity (\ref{specdef}) is valid
only up  to the $x$-independent constant. Let us denote by
$y_{j,\sigma}(x)$ the eigenvalues which are asymptotic to \be
y_{j,\sigma}(x) = \frac { t_{rn_j,j}}{n_j(x-\gamma)^{r+1} } +
{\omega_j}^\sigma \frac {t_{rn_j-1,j}}{n_j(x-\gamma)^{r+1-\frac 1
    {n_j}}} + \dots\ ;\ \ \ \sigma  = 0,\dots,n_j-1\label{evalll}
\ee
Note that these $n_j$ eigenvalues are cyclically permuted by a small
loop around the singularity $\gamma$ on the base curve. Moreover
suppose now that there are two such cyclic multiplets with the same
order $n_j$  and same leading term $t_{rn_j}$ (which means that the
leading term at the singularity of $A(x)$ has two Jordan blocks of the
same size and the same eigenvalue); then our genericity assumption
(\ref{de:generic}) precisely implies that the next-to-leading
coefficients displayed in (\ref{evalll}) will distinguish the two
cyclic groups. We denote accordingly the corresponding spectral
projectors
\be
\Pi_{j,\sigma}(x) := \frac {\widetilde {(A-y_{j,\sigma}})(x)}{\tr (\widetilde {A-y_{j,\sigma}})(x)}\ .
\ee
Then the previous deformation matrices can be written as (restoring the label $\gamma$)
\bea
\mathcal T^\gamma_{J,j} &\!\!\!\!=& \frac 1
J\le(\sum_{\sigma=0}^{n_j^\gamma-1} (x-\gamma)^{-\frac J{n_j^\gamma}}
{\omega_{\gamma,j}}^{J\sigma} \Pi_{j,\sigma} \ri)_{pp} ,\ \
\omega_{\gamma,j}:={\rm e}^{2i\pi/n_j^\gamma}\\
\mathcal C _\gamma  &\!\!\!\!=&  A_\gamma(x)\ .
\eea
Note that the expressions here above which may in principle contain
fractional powers of the local parameter, in fact do not because of
the cyclicity properties.

Equation (\ref{evalll}) (and an analogue for $x=\infty$) could be
rephrased in more geometrical terms by saying that the  desingularized
spectral curve $\Sigma$  above each point $x=\gamma$ has $s$(=number
of Jordan cells) distinct points $\zeta_\gamma^j$  which are all
branchpoins of order $n_j^\gamma$  for the map $x:\Sigma \to \C P^1$,
for which a local parameter is $q_{\gamma,j}
:=(x-\gamma)^{1/n_j^\gamma}$ (or $q_{\infty,j}:= x^{-1/n_j^\infty}$).
With these notations in place we can finally prove
\bp[Tau function for minimally resonant irregular/Fuchsian isomonodromic deformations]
The following differential is closed and hence is the differential of a locally defined function
\be
{\rm d}\ln \tau := \sum_{j=0}^{s_\infty} \sum_{J=1}^{n_j^\infty}
\frac{1}J {\rm d}t_{J,j}^\infty \res{\zeta_\infty^j} x^{J/{n_j^\infty}}  y{\rm d}x +
\sum_{\gamma\in \mathcal D} \sum_{j=0}^{s_\gamma}
\sum_{J=1}^{n_j^\gamma} {\rm d}t_{J,j}^{\gamma} \frac {1} {J}
\res{\zeta_\gamma^j} \frac {y{\rm d}x}{ (x-\gamma) ^{-J/{n_j^\gamma
}}}  + \frac 1 2 \sum_{\gamma\in \mathcal D} \res{x=\gamma} \tr (A^2)
    {\rm d}\gamma
\ee
Here the residues of the differential $y{\rm d}x$ are taken at points
of the spectral curve (i.e. choosing the appropriate eigenvalue
$y_{\gamma,\sigma}$) and are taken with respect to the local
parameters by going around the point $x=\gamma$ of the base-curve
$n_k^\gamma$ times.

Moreover the isomonodromic times and the parameters of formal monodromy for the irregular singularities are obtained by the residues
\bea
t_{J,j}^{\infty} & =&\res{\zeta_\infty^j} x^{-J/n_j^\infty} y{\rm d}x \ , \ \  \  j=1,\dots s_\infty,\  J = 1,\dots, n_j^\infty \\
t_{J,j}^{\gamma} &=& \res{\zeta_\gamma^j} (x-\gamma)^{J/n_j^\gamma} y{\rm d}x \ ,\ \ \ j=1,\dots,s_\gamma,\ J=1,\dots, n_j^\gamma,\ \infty\neq \gamma \in \mathcal D\\
t_{0,j}^{\gamma} &=& \res{\zeta_\gamma^j} y{\rm d}x + \frac {n_j^\gamma-1}2\ ,\ \
\eea
\ep
{\bf Proof}.
The residue-formul\ae \ for the isomonodromic times follow from the expansion of the eigenvalues in the respective local parameter. The  parameter of formal monodromy $t_{0,\gamma}$ have a correction which was explained in (\ref{strange}) and follows simply from
\be
\res{\gamma} \tr{D^{\gamma}_j(x)}{\rm d}x= \res{\zeta_\gamma^j} y{\rm d}x  =t_{0,j}^{\gamma}-\frac {n_j^\gamma-1}2\ ,
\ee
where $D_{j,\gamma}$ was introduced in (\ref{diagbare}).
The proof uses the same arguments adopted earlier but it
is only more involved due to  the presence of multiple singularities
and times.
First of all we note that the residues can be pushed down on the base-curve
\be
\frac{1}J  \res{\zeta_\infty^j} (x-\gamma)^{-J/{n_j^\gamma}}  y{\rm d}x =\frac 1 J\res{x=\gamma}\le( \sum_{\ell=0}^{n_j^\gamma} (x-\gamma)^{-J/{n_j^\gamma}} (\omega_{\gamma,k})^{J\sigma} y_{\gamma,k,\sigma} \ri){\rm d}x
\ee
because the sum in the bracket is a {\em bona fide} (local) function of $x$ without fractional powers, due to the periodicity of the cyclic $n_j^\gamma$-plet of eigenvalues. Here the symbol $y_{\gamma,k,\sigma}$ denotes the $\sigma$-member of the $k$-th  multiplet (corresponding to the $k$-th block in the Jordan-cell decomposition) of eigenvalues near $x=\gamma$.
 We compute the closure of the differential
\bea JK \pa_{t_{J,j}^\gamma}\pa_{t_{K,k}^\mu}\ln\tau &=&
J\res{x=\mu} \le(\sum_{\sigma=0}^{n_k^\mu-1}
(\omega_{\mu,k})^{K\sigma}(x-\mu)^{-\frac{K}{n_k^\mu}}
\pa_{t_{J,j}^\gamma}y_{\mu, k,\sigma}\ri) {\rm d}x= \nonumber \eea
\bea &=& J\res{x=\gamma} \le( \sum_{\sigma=0}^{n_k^\mu-1}
(\omega_{\mu,k})^{K\sigma}(x-\mu)^{-\frac K{n_k^\mu}} \tr\le(\frac
{\widetilde{A-y_{\mu,k,\sigma}}}{\tr
(\widetilde{A-y_{\mu,k,\sigma}})}
\frac{\pa A}{\pa t_{J,j}^\gamma} \ri) \ri){\rm d}x=  \nonumber \\
&=& J\res{x=\gamma}\le(  \sum_{\sigma=0}^{n_k^\mu-1}
     {\omega_{\mu,k}}^{K\sigma}(x-\mu)^{-\frac K{n_k^\mu}} \tr\le(
     \Pi_{\mu,k,\sigma}(x)
\le(\frac {{\rm d}\mathcal T_{J,j}^{\gamma} }{{\rm d}x} - [A,\mathcal
  T_{J,j}^{\gamma}]\ri)\ri) \ri){\rm d}x= \nonumber \\
&=& \res{x=\mu}  \sum_{\sigma=0}^{n_k-1}
   {\omega_{\mu,k}}^{K\sigma}(x-\mu)^{-\frac K{n_k^\mu}}
   \tr\le(\Pi_{\mu,k,\sigma} (x) \frac {\rm d}{{\rm
       d}x}\le(\sum_{\rho=0}^{n_j^\gamma-1} (x-\gamma)^{-\frac
     J{n_j^\gamma}} {\omega_j}^{J\rho} \Pi_{\gamma,j,\rho} \ri)_{pp}
   \ri) {\rm d}q =\nonumber\\
  &=&\res{x=\mu}
  \res{z=\gamma}\le( \sum_{\sigma=0}^{n_k^\mu-1}\sum_{\rho=0}^{n_j^\gamma
    -1} {\omega_{j,\mu}}^{J\rho}
      {\omega_{k,\mu}}^{K\sigma}(z-\gamma)^{-\frac J{n_j^\gamma}}
      (x-\mu)^{-\frac K{n_k^\mu}} \frac{\tr\le(
  \Pi_{\mu,k,\sigma}(x)   \Pi_{\gamma,j,\rho}(z)  \ri)}{(z-x)^2}\ri) {\rm
    d}z{\rm d}x \nonumber
\eea
This formula is symmetric because we can exchange the order of the
residues; indeed if $\mu\not= \gamma$ the order of residues is
certainly irrelevant, whereas if $\mu=\gamma$ we can exchange the
order because the residue of the bidifferential vanishes due to Lemma
\ref{bidiff}.  In these computations we have assumed that both
$\gamma,\mu$ are finite poles, but the proof goes through similarly
with minor modification only in the notation and local parameter if
one or both coincide with the pole at $\infty$ and if one or both the
deformations are translations of the position of the (finite)
pole. Q.E.D.\par \vskip 5pt
As an immediate corollary of the proof we have
\bc[Hessian of the Tau function]
The second derivatives of the Tau function are expressed in terms of the spectral curve according to the formul\ae (for the points at $\infty$ the formula needs obvious modifications).
\bea
\frac {\pa^2 \ln \tau}{\pa t_{J,j}^\gamma \pa t_{K,k}^{\mu}} =\res{x=\mu}
  \res{z=\gamma}\le( \sum_{\sigma=0}^{n_k^\mu-1}\sum_{\rho=0}^{n_j^\gamma
    -1} {\omega_{j,\mu}}^{J\rho}
      {\omega_{k,\mu}}^{K\sigma}(z-\gamma)^{-\frac J{n_j^\gamma}}
      (x-\mu)^{-\frac K{n_k^\mu}} \frac{\tr\le(
  \Pi_{\mu,k,\sigma}(x)   \Pi_{\gamma,j,\rho}(z)  \ri)}{(z-x)^2}\ri) {\rm
    d}z{\rm d}x \nonumber
\eea

\ec
\appendix
\section{An example: Airy-like equations}
In this appendix we provide an example of isomonodromic deformation of
an irregular singularity which is of almost trivial nature but
illustrates the machinery developed earlier. We consider an arbitrary
polynomial $V(x)$ of degree $n+1$ and the following scalar ODE (note
that the simplest nontrivial case is Airy's equation)
\bea
V'(\pa_x) f = x f(x)\\
V(z)=\sum_{j=1}^{n+1} \frac{v_j}j z^j
\eea
The matrix first order ODE associated to this equation is
\be
\Psi' := \le[
\begin{array}{cccc}
 & 1 &&\\
&&\ddots&\\
&&&1\\
\ds \frac{x-v_1}{v_{n+1}}& \ds \frac{-v_2}{v_{n+1}} & \dots & \ds \frac {-v_{n}}{v_{n+1}}
\end{array}
\ri]\Psi
\ee
The matrix $A(x)$ is in fact the companion matrix of the polynomial
$V'(y)$ and thus the spectral curve is simply the rational curve
(genus zero)
\be
V'(y)=x\ .
\ee
The only singularity is at $x=\infty$ and it is irregular; the leading
coefficient is nilpotent with canonical form $0^2(0)^{d-2}$. In a
certain sense this is a {\em non-example} of our setting because the
Lidskii submatrix of the subleading term does not satisfy our
genericity assumption. However it is sufficient to perform a single
shearing gauge transformation to recast the problem to a non-resonant
one and  this feature is the only essential ingredient to
carry out the analysis as in Sect. \ref{IsoDef}.
Following the same steps as in Prop. \ref{lemma2} (with the only
difference that now we have to perform the inverse shearing), we
obtain that there exists a formal solution of the form ($q=x^{\frac
  1n}$, $t_{n+1} := n(v_{n+1})^{-1/n}$)
\bea
\Psi(x) &=& {(t_{n+1}/n)}^{-G} Y(x) {\rm e}^{Q(x)}x^{\frac {t_0+G}n} =
    {(t_{n+1}/n)}^{-G} q^G Z(q) {\rm e}^{T(q)}W^{-1}\ ,\qquad \ G:={\rm  diag}(0,1,\dots,n-1) \nonumber\\
&&Y(x) := \sum_{j=0}^\infty \le[
\begin{array}{cccc}
&&&x^{-1}\\
1&&&\\
&\ddots&&\\
&&1&
\end{array}
\ri]^j z_j
\qquad Q(x):= \sum_{j=1}^{n+1} \frac {t_j}j \le[\begin{array}{cccc}
&1&&\\
&&\ddots&\\
&&&1\\
x&&&
\end{array}
\ri]^j  =: \sum_{j=1}^{n+1} \frac {t_j}j \mathcal H^j(x)\nonumber  \\
&& Z(q) := \sum_{j=0}^{\infty} q^{-j} \mathcal C^{-j} z_j W\ ,\ \
T(q):= \sum_{j=1}^{n+1} \frac {t_j}j q^j \Omega^{-j} + t_0\ln (q)\nonumber\\
&& \mathcal C:= \le[\begin{array}{cccc}
&1&&\\
&&\ddots &\\
&&&1\\
1&&&
\end{array}\ri]\ ,\qquad W:=[\omega^{-(i-1)(j-1)}]_{i,j}\ ,\ \
\omega:= {\rm e}^{\frac {2i\pi}n}\nonumber\\
&&z_0:= \1\ ,\ \ z_j=\hbox{ diagonal matrices}.
\eea
Note that the asymptotic representation is unique and is an explicit
function of the parameters $t_j$ because the matrix $A$ is completely
determined by them and the diagonal matrices  $z_j$ are constructed by a determined recursive
procedure along the lines of Prop. \ref{lemma2}.
Of the two asymptotic presentations the one in fractional powers is
the usual one in the context of Airy-like equations, whereas the one
in terms of integer powers is the one we have used in the present paper.
The coefficients $t_j$ are the expansion of the eigenvalue in
fractional powers
\bea
y(x)&=& \frac 1 n\sum_{j=1}^{n+1} t_j x^{\frac {j}n -1 }\nonumber\\
t_j &+&\delta_{j0}\frac {n-1}2=  -\res{x=\infty} x^{-j/n} y{\rm
d}x \nonumber \\&=& -\res{y=\infty} (V'(y))^{-\frac jn} y
V''(y){\rm d}y = \le\{
\begin{array}{lc}
\ds \frac {n}{n-j}
\res{y=\infty} (V'(y))^{1-\frac jn}{\rm d}y & ,j<n\\
\ds t_n=-\frac{v_n}{v_{n+1}}&\\
\ds t_{n+1} = n(v_{n+1})^{-1/n} &
\end{array}
\ri.
\eea
note that $t_0=\frac {1-n}2$ because the curve is of  genus zero, indeed
\be
t_0+ \frac {n-1}2 = -\res{x=\infty} \tr D_{bare}(x){\rm d} x= -\res{x=\infty} A(x){\rm d}x=0\ .
\ee
where $D_{bare}$ is the bare form of the differential equation (see (\ref{bareairy})); the residues of the traces of $D_{bare}$ and $A$ coincide because they are connected by a formally analytically invertible gauge giving a $\mathcal O(x^{-2})$ extra term which does not contribute to the residue.

One can write an explicit integral representation of the
solutions in terms of Fourier--Laplace integrals
\be
f_k(x):= \int_{\Gamma_k} {\rm e}^{xy-V(y)}\ \Rightarrow\
\Psi_{j,k}(x)= \int_{\Gamma_k} y^j  {\rm e}^{xy-V(y)} = \pa_x^j f_k(x)\label{airy}
\ee
where the contours $\Gamma_k$ can be chosen in $n$ ``homologically''
independent ways \cite{BEH2}: in the case $n=2$ one recognizes the standard integral representation of Airy's functions. The formal asymptotic representation in
fractional powers as well as the Stokes matrices can be obtained by
the steepest descent method, whereas the integer powers asymptotics
cannot.
The bare Pfaffian system is given by
\bea
\le(\pa_x - \sum_{j=1}^{n+1}\frac {t_j}{nx} \mathcal H^j - \frac
G{nx}\ri)\Psi_{bare}=0\ ,\qquad \label{bareairy}
\le(\pa_{t_j} - \frac 1{j} \mathcal H^j(x)\ri)\Psi_{bare}=0\\
\Psi_{bare}:= {\rm e}^{Q(x)} x^{\frac Gn}
\eea
\subsection{Isomonodromic deformations}
In this case there is no monodromy in the usual sense, only Stokes'
matrices, and these are the preserved data under the deformation.
The parameters of deformations are the $t_j$ or -which is the same
after a change of coordinates- the coefficients $v_j$ of the ``potential''
(this terminology comes from the application to random matrices).
However our connection $A(x)$ is not gauge--fixed, but it could be
done so by conjugating it by the constant coefficient of the series
$Y(x)$ (which is an explicit function of the $t_j$'s).

The deformation equations are easy to describe because we have an
explicit solution. Indeed it is immediate from the integral
representation (\ref{airy}) that
\be
\pa_{v_j} \Psi = -\frac  1 j\pa_x^{j} \Psi
\ee
and hence the matrices
\be
\mathcal T_j := -\frac  1 j\pa_x^j \Psi\, \Psi^{-1}
\ee
trivially satisfy the zero curvature conditions; note that they are
polynomials of degree at most $1$ except for $T_{n+1}$ which is of
degree $2$.  The deformation
equations in terms of the parameters $t_j$ can be obtained by using
the Jacobian of the change of coordinate from $v_j$ to $t_j$.
Since these deformations are for a connection non gauge-fixed
 the deformation equations are in the more general
form of Thm. \ref{thm:isogen}.

\subsection{Tau function}
Since the spectral curve is rational, the tau function actually
coincides with the tau function of the Whitham hierarchy
\bea
{\rm d}\ln \tau &=& \sum_{j=1}^{n+1} H_j{\rm d}t_j \\
H_j&:=& - \frac  1 j \res{x=\infty} x^{j/n} y{\rm d}x = \frac
n{j(n+j)} \res{y=\infty} (V'(y))^{1+\frac j n} {\rm d}y \eea It is
an exercise left to the reader to check that an integral is \be
\ln \tau = \frac 1 2 \sum_{j=1}^{n-1} t_j H_j \frac {n+1-j}{n+1}
\ee For the case of (translated/dilated/gauged) Airy's equation
($n=2$) \be v_3 f'' + v_2 f' + v_1 f = x f \ee the tau function is
\be \ln \tau_{Ai} = -\frac {2{t_1}^3}{3 t_3} = \frac
    {(4v_1v_3-v_2^2)^3}{768 {v_3}^4}
\ee
In all cases $n>2$  it is always a rational expression in the
parameters $t_j$ or $v_j$. Note that this equation could be transformed to the standard Airy equation by a scalar gauge transformation, a translation and a dilation, which would eliminate all the parameters. The first really nontrivial case is the "hyper"-Airy equation
\bea
v_4 f''' + v_3 f'' + v_2 f' + v_1 f = x f\ ,
\eea
for which
\bea
&&\ln \tau_{hAi} = -\frac {3 t_2(12 {t_1}^2 t_4 + {t_2}^3)} {8 {t_4}^2}\ .
\eea

\end{document}